\documentclass[11pt]{article}

\usepackage[T1]{fontenc}
\usepackage{jheppub}
\usepackage{wasysym}
\usepackage{color}

\usepackage{tabularx}
\usepackage{physics}
\usepackage{tikz}
\usetikzlibrary{calc}
\usepackage{graphicx} 
\usepackage{ae}
\usepackage{comment}

\usepackage{amsfonts,amsmath,amsthm,amssymb,amsbsy}
\usepackage{mathtools}

\usepackage{pifont}
\newcommand \roundrectangle[6] {\draw[rounded corners=#1 pt,thick,rotate=#6] ($(#2-#4*0.5,#3-#5*0.5)$) rectangle ($(#2+#4*0.5,#3+#5*0.5)$);}
\newcommand \fillroundrectangle[6] {\filldraw[gray!20,rounded corners=#1 pt,thick,rotate=#6] ($(#2-#4*0.5,#3-#5*0.5)$) rectangle ($(#2+#4*0.5,#3+#5*0.5)$);}

\title{Amplitubes: Graph Cosmohedra}

\author[]{Ross Glew and}\emailAdd{r.glew@herts.ac.uk}
\author[]{Tomasz \L ukowski}\emailAdd{t.lukowski@herts.ac.uk}

\affiliation[]{Department of Physics, Astronomy and Mathematics, \\ University of Hertfordshire, \\  Hatfield, Hertfordshire, AL10 9AB, United Kingdom}

\abstract{The tree-level scattering amplitudes for $\tr(\phi^3)$ theory can be interpreted as a sum over the vertices of a polytope known as the associahedron. For each graph $G$, there exists a natural generalisation of the associahedron, which is constructed by considering tubes and tubings of the underling graph. This family of polytopes are called {\it graph associahedra}. The classical associahedra then arise as the graph associahedron for the path graphs. It is therefore natural to associate to each graph associahedron an amplitude-like object, we refer to as the {\it amplitube}, defined via a sum over its vertices. Recently, also in the context of $\tr(\phi^3)$ theory, progress has been made towards defining a new geometric object, coined the {\it cosmohedron}, which computes not the amplitude, but the cosmological wavefunction as a sum over its vertices. This polytope can be constructed by consistently blowing up all boundaries of the associahedron to co-dimension one. Building on these results, in the present paper, we generalise the notion of the wavefunction for arbitrary graphs. These new expressions, which we call {\it cosmological amplitubes}, are defined via a sum over the vertices of a corresponding polytope, the {\it graph cosmohedron}. The graph cosmohedra are constructed by considering {\it regions} and {\it regional tubings} of the underlying graph which we introduce. Like the cosmohedron, the graph cosmohedra can be obtained by consistently blowing up all boundaries of the corresponding graph associahedron to co-dimension one. This new family of polytopes constitutes a vast generalisation of the cosmohedron, and we provide explicit embeddings for them, which builds upon an ABHY-like embedding for the graph associahedra.}

\begin{document}

\maketitle
\pagebreak

\section{Introduction}
In recent years the study of scattering amplitudes has fostered a rich interplay between mathematics and physics. As research progresses, it becomes increasingly clear that much of the structure of scattering amplitudes is governed by purely geometrical, or perhaps even combinatorial, principles. In fact, the observation that cubic Feynman diagrams can be used as labels of the vertices of a polytope, the associahedron, provides the first hint of a possible geometrical interpretation of amplitudes. In the amplitudes inspired description of the associahedron each facet corresponds to a {\it factorisation} channel of the $n$-point amplitude, and each vertex corresponds to a cubic Feynman graph. The tree-level scattering amplitudes of $\tr(\phi^3)$ theory for example can then be viewed as a sum over vertices of the associahedron. In this context the familiar statements of {\it locality} and {\it unitarity} of the amplitude translate into statements about factorisation properties of boundaries of the associahedron. By taking the planar dual of the tree-level Feynman diagram this can be phrased in terms of triangulations of an $n$-gon. In this way each boundary of the associahedron is labelled by a partial triangulation of an $n$-gon, with the vertices corresponding to full triangulations, and the facets corresponding to single chords. The sum over vertices of the associahedron then becomes 
\begin{align}
A_n = \sum_{T} \prod_{(ij) \in T} \frac{1}{X_{ij}},
\end{align}
where the sum is over all triangulations of the $n$-gon and $X_{ij}$ correspond to the chords of the triangulation. Upon identifying the $X_{ij}$ with the planar Mandelstam variables, which are the squares of sums of consecutive momenta, the above expression recovers the tree-level amplitudes of $\tr(\phi^3)$ theory. This combinatorial statement was made concrete by the discovery of the ABHY associahedron \cite{Arkani-Hamed:2017mur}, a positive geometry \cite{Arkani-Hamed:2017tmz} that provides a realisation of the associahedron in the kinematic space of $n$-point massless scattering whose canonical form encodes tree-level amplitudes in $\tr(\phi)^3$ theory.

More recently techniques developed in the study of scattering amplitudes have started to be applied in a cosmological setting, see \cite{Arkani-Hamed:2024jbp,Arkani-Hamed:2023kig,Arkani-Hamed:2023bsv,Arkani-Hamed:2017fdk,De:2023xue,Benincasa:2024leu,Benincasa:2024lxe,De:2024zic} and references therein for advances in this direction. The object of study in this context are the cosmological wavefunctions. At the level of combinatorics the wavefunction is described not by triangulations of an $n$-gon but rather by {\it nested polyangulations} or {\it Russian dolls} \cite{Arkani-Hamed:2024jbp}. In terms of nested polyangulations the wavefunction takes the following form
\begin{align}
\Psi_{n} = \sum_{{\bf P}} \prod_{P \in {\bf P}} \frac{1}{\mathcal{P}_P}.
\end{align}
Here the sum is over all maximal sets ${\bf P}$ of non-overlapping\footnote{By non-overlapping we mean that the chords defining each sub-polygon are non-overlapping.} sub-polygons of the $n$-gon, and the $\mathcal{P}_{P}$ are variables associated to the perimeter of each sub-polygon $P$. Much like the associahedron, the nested polyangulations can be used to define their own polytope named the {\it cosmohedron} \cite{Arkani-Hamed:2024jbp}. The facets of the cosmohedron correspond to all partial triangulations of the $n$-gon, and as such are in bijection with the boundaries of the asociahedron, whereas the vertices correspond to maximally nested polyangulations.
 
Meanwhile, in the mathematics literature, the above developments have lead to the study of amplitude-like expressions which satisfy their own versions of locality and unitarity. These include for instance the CEGM amplitudes of \cite{Cachazo:2019ngv} which provide a grassmannian generalisation of the $\tr(\phi^3)$ amplitudes. Other examples of amplitude-like constructions include those provided by matroids \cite{Lam:2024jly} and the surface functions of \cite{Arkani-Hamed:2024pzc}. To motivate the main topic of study in this paper, we consider yet another interpretation of the boundary stratification of the associahedron in terms of {\it tubes} and {\it tubings} of a path graph on $(n-3)$ vertices. A tube of a graph is defined as a subset of vertices of the graph which induce a connected subgraph. Whilst a tubing is a collection of tubes which are either nested or do not intersect and are not adjacent on the graph. In this language the facets of the associahedron are labelled by single tubes of the path graph, and the vertices are labelled by maximal tubings. The advantage of this definition is that it can immediately be applied to arbitrary graphs in which case the associahedron is replaced with the larger set of polytopes named {\it graph associahedra} \cite{carr2006coxeter}. Since the amplitude associated to the associahedron is defined via a sum over its vertices it is natural to extend this definition to arbitrary graph associahedra, where we refer to the corresponding functions as {\it amplitubes} defined as 
\begin{align}
 A_G = \sum_{\tau\in\Gamma_G^{\text{max}}} \prod_{t \in \tau} \frac{1}{X_t},
 \end{align} 
where $\tau$ sums over all maximal tubings of the graph $G$, and $X_t$ are formal variables associated to each tube $t$. In the case where we take $G$ to be the path graph the above formula reduces to the familiar $\tr(\phi^3)$ amplitudes. For the amplitube, the analogue of locality is that the variables appearing in the denominator of $A_G$ all correspond to {\it connected} subgraphs, and unitarity states that upon taking a residue at $X_t=0$ the amplitube factorises into a product of two simpler amplitubes. Furthermore, there exists an ABHY-like embedding of the graph associahedra in the {\it graph kinematic space} spanned by the set of $X_t$. By calculating the canonical form of the graph associahedron and pulling-back to an appropriate subspace one recovers the associated amplitube. 

The focus of this paper will be to extend the notion of amplitubes into a cosmological setting by generalising nested polyangulations into tubing notions which can then be applied to arbitrary graphs in order to define {\it cosmological amplitubes}. Guided by the results for $\tr (\phi^3)$ theory, we will see that the cosmological amplitubes are naturally defined in terms of objects we refer to as {\it regional tubes} and {\it regional tubings} as 
\begin{align}
\Psi_{G} = \sum_{\rho\in \Phi_G^{\text{max}}} \prod_{r \in\rho} \frac{1}{\mathcal{R}_r}.
\end{align}
Here we sum over all maximal regional tubings $\rho$ and take the product over all regions $r$ of $\rho$, which play the role of nested polyangulations and sub-polygons, respectively, for arbitrary graphs. Just as for the $\tr (\phi^3)$ wavefunctions, these functions also have their own geometric counterpart which we refer to as {\it graph cosmohedra} whose boundary structure encode the cosmological amplitubes. As such each facet of the graph cosmohedron will correspond to a boundary of the corresponding graph associahedron. As we shall explain the graph cosmohedra can be realised in the graph kinematic space by building upon the ABHY-like embedding of the graph associahedra. In the case of the empty graph, path graph and fully connected graph we show that graph cosmohedra reproduce the permutohedron, cosmohedron and permutoassociahedron respectively. In this way the graph cosmohedra can be seen as an interpolation between the permutohedron and the permutoassociahedron \cite{kapranov1993permutoassociahedron}. This suggests that, from a mathematical perspective, the graph cosmohedra constructed here might more naturally be referred to as {\it graph permutoassociahedra}.

This paper is organised as follows. In Section \ref{sec:graph_ass} we introduce the notions of tubes and tubings of graphs which are needed to define the amplitubes and their associated graph associahedra. Many of the properties of the graph associahedra have already appeared in the literature and we will spend the majority of the first part of the paper collecting and rephrasing these results in term of tubings. Most importantly this will include an ABHY-like embedding, we refer to as the {\it tubing embedding}, for general graph associahedra which can be simply stated when phrased in terms of tubes. In Section \ref{sec:graph_cosmo} we introduce the notion of regions and regional tubings of arbitrary graphs which are needed in order to define the cosmological amplitubes and their corresponding graph cosmohedra. Having defined the combinatorial object of interest, we turn to providing a geometric realisation of the graph cosmohedra. We provide an explicit embedding of the graph cosmohedra in the graph kinematic space and present explicit examples of graph cosmohedra for the empty, path and complete graph. At the end of this section we discuss the factorisation properties of graph cosmohedra on the co-dimension one facets. In Section \ref{sec:outlook} we conclude and comment on possible future research directions.

\section{Graph Associahedra}\label{sec:graph_ass}
We start in this section by recalling the definition and basic facts about tubes and tubings of a graph. Our discussion will follow closely the definitions provided in \cite{carr2006coxeter}. These definitions will allow us to define the corresponding amplitube of a graph. After the combinatorial construction, we provide an embedding of the graph associahedra, which we refer as the {\it tube embedding}, that will be the starting point for the definition of graph cosmohedra in the next section. The tube embedding for graph associahedra we provide combines the original ABHY embedding of the associahedron \cite{Arkani-Hamed:2017mur} with the embedding in \cite{devadoss2009realization} for general graph associahedra.

\subsection{Tubes and Tubings}
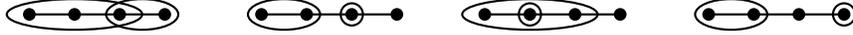
\begin{figure}
\center
 \begin{tikzpicture}[scale=1.2]
        \coordinate (A) at (0,0);
        \coordinate (B) at (1/2,0);
        \coordinate (C) at (1,0);
        \coordinate (D) at (3/2,0);
        \draw[thick, black] (1.25,0) ellipse (0.4cm and 0.17cm);
        \draw[thick, black] (0.5,0) ellipse (0.75cm and 0.17cm);
        \draw[thick] (A) -- (B) -- (C) -- (D);
        \fill[black] (A) circle (2pt);
        \fill[black] (B) circle (2pt);
        \fill[black] (C) circle (2pt);
        \fill[black] (D) circle (2pt);
    \end{tikzpicture} \quad \quad 
 \begin{tikzpicture}[scale=1.2]
        \coordinate (A) at (0,0);
        \coordinate (B) at (1/2,0);
        \coordinate (C) at (1,0);
        \coordinate (D) at (3/2,0);
        \draw[thick, black] (0.25,0) ellipse (0.4cm and 0.17cm);
        \draw[thick] (A) -- (B) -- (C) -- (D);
        \fill[black] (A) circle (2pt);
        \fill[black] (B) circle (2pt);
        \fill[black] (C) circle (2pt);
        \fill[black] (D) circle (2pt);
        \draw[black,thick] (C) circle (3.5pt);
 \end{tikzpicture}\quad \quad 
\begin{tikzpicture}[scale=1.2]
        \coordinate (A) at (0,0);
        \coordinate (B) at (1/2,0);
        \coordinate (C) at (1,0);
        \coordinate (D) at (3/2,0);
        \draw[thick, black] (0.5,0) ellipse (0.75cm and 0.17cm);
        \draw[thick] (A) -- (B) -- (C) -- (D);
        \fill[black] (A) circle (2pt);
        \fill[black] (B) circle (2pt);
        \fill[black] (C) circle (2pt);
        \fill[black] (D) circle (2pt);
        \draw[black,thick] (B) circle (3.5pt);
\end{tikzpicture} \quad \quad
\begin{tikzpicture}[scale=1.2]
        \coordinate (A) at (0,0);
        \coordinate (B) at (1/2,0);
        \coordinate (C) at (1,0);
        \coordinate (D) at (3/2,0);
        \draw[thick, black] (0.25,0) ellipse (0.4cm and 0.17cm);
        \draw[thick] (A) -- (B) -- (C) -- (D);
        \fill[black] (A) circle (2pt);
        \fill[black] (B) circle (2pt);
        \fill[black] (C) circle (2pt);
        \fill[black] (D) circle (2pt);
        \draw[black,thick] (D) circle (3.5pt);
    \end{tikzpicture}
    \caption{An illustration of the four possible configurations of tubings: intersecting, adjacent, nested and non-adjacent non-intersecting tubes. The last two pairs of tubes are compatible. }
    \label{fig:tub_config}
\end{figure}

Let $G$ be a graph with the vertex set $V_G$ and edge set $E_G$. A {\it tube} $t=\{v_1,v_2,\ldots,v_{|t|}\}\subset V_G$ on $G$ is a proper non-empty subset of vertices of $G$ such that the induced subgraph $G[t]$ is connected. The set of all tubes of $G$ is denoted by $T_G$. We also define $\overline{T}_G=T_G\cup \{V_G\}$, and in this context, we call the graph vertex set $V_G$ {\it the root}. We introduce the following terminology relating two tubes $t_1$ and $t_2$ on a given graph: we say that
\begin{itemize}
\item $t_1$ and $t_2$ {\it intersect} if $t_1 \cap t_2 \neq \emptyset$ and $t_1 \not\subset t_2$ and $t_2 \not\subset t_1$,
\item $t_1$ and $t_2$ are {\it adjacent} if $t_1 \cap t_2 = \emptyset $ and $ t_1 \cup t_2 \in \overline{T}_G$,
\item $t_1$ is {\it nested} in $t_2$ if $t_1$ is a proper subset of $t_2$: $t_1 \subsetneq t_2 $,
\item $t_1$ and $t_2$ are {\it compatible} if they do not intersect and they are not adjacent.
\end{itemize}
These concepts are easily visualised as displayed in Fig.~\ref{fig:tub_config}.

A subset of tubes $\tau=\{t_1,t_2,\ldots,t_{|\tau|}\} \subset T_G$ is called a {\it tubing} of $G$ if all tubes in $\tau$ are mutually compatible and the set $V_G\setminus \bigcup_i t_i$ is not empty. A tubing is said to be {\it maximal} if no more compatible tubes can be added. We denote the set of all tubings of $G$ by $\Gamma_G$ and the set of all maximal tubings by $\Gamma^{\text{max}}_G$. Given a tubing $\tau$ of $G$, we define $\overline{\tau} =\tau \cup V_G$. Then we can define a graded poset $\mathcal{P}_{\overline{\tau}}(G)$ on the set of tubes $t \in \overline{\tau}$ by inclusion, where the rank function is given by the number of tubes in $\tau$ between a given tube and the root. Examples of tubings $\tau$ and their corresponding posets $\mathcal{P}_{\overline{\tau}}(G)$ are 
\begin{align}
\mathcal{P} 
\left(
\tikz[baseline,yshift=0.5ex] \node[anchor=base] at (0,0) {\begin{tikzpicture}[scale=0.8]
        \coordinate (A) at (0,0);
        \coordinate (B) at (1/2,0);
        \coordinate (C) at (1,0);
        \coordinate (D) at (3/2,0);
        \draw[thick, black] (0.25,0) ellipse (0.4cm and 0.17cm);
        \draw[thick] (A) -- (B) -- (C) -- (D);
        \roundrectangle{0}{0.75}{0}{2.2}{0.6}{0};
        \fill[black] (A) circle (2pt);
        \fill[black] (B) circle (2pt);
        \fill[black] (C) circle (2pt);
        \fill[black] (D) circle (2pt);
        \draw[black,thick] (D) circle (3.5pt);
    \end{tikzpicture}};
\right) = \tikz[baseline,yshift=2.5ex] \node[anchor=base] at (0,0) {\begin{tikzpicture}[scale=0.8]
        \coordinate (A) at (0,0);
        \coordinate (B) at (-1/4,-1/2);
         \coordinate (C) at (+1/4,-1/2);
         \coordinate (D) at (-1/4,-1);
        \draw[thick] (A) -- (B);
        \draw[thick] (A) -- (C);
        \node[fill=black, shape=rectangle, minimum size=1.5mm, inner sep=0mm] at (A) {};
        \fill[black] (B) circle (2pt);
        \fill[black] (C) circle (2pt);
    \end{tikzpicture}};, \ 
    \mathcal{P} 
\left(
\tikz[baseline,yshift=0.5ex] \node[anchor=base] at (0,0) {\begin{tikzpicture}[scale=0.8]
        \coordinate (A) at (0,0);
        \coordinate (B) at (1/2,0);
        \coordinate (C) at (1,0);
        \coordinate (D) at (3/2,0);
        \draw[thick, black] (0.5,0) ellipse (0.8cm and 0.25cm);
        \roundrectangle{0}{0.75}{0}{2.2}{0.62}{0};
        \draw[thick] (A) circle (3.5pt);
                \draw[thick] (C) circle (3.5pt);
        \draw[thick] (A) -- (B) -- (C) -- (D);
        \fill[black] (A) circle (2pt);
        \fill[black] (B) circle (2pt);
        \fill[black] (C) circle (2pt);
        \fill[black] (D) circle (2pt);
    \end{tikzpicture}};
\right) &= \tikz[baseline,yshift=3.6ex] \node[anchor=base] at (0,0) {\begin{tikzpicture}[scale=0.8]
        \coordinate (A) at (0,0);
        \coordinate (B) at (0,-1/2);
         \coordinate (C) at (-1/2,-1);
         \coordinate (D) at (1/2,-1);
        \draw[thick] (A) -- (B) -- (C);
                \draw[thick]  (B) -- (D);
        \node[fill=black, shape=rectangle, minimum size=1.5mm, inner sep=0mm] at (A) {};
        \fill[black] (B) circle (2pt);
        \fill[black] (C) circle (2pt);
                \fill[black] (D) circle (2pt);
    \end{tikzpicture}};, \ 
    \mathcal{P} 
\left(
\tikz[baseline,yshift=0.5ex] \node[anchor=base] at (0,0) {\begin{tikzpicture}[scale=0.8]
        \coordinate (A) at (0,0);
        \coordinate (B) at (1/2,0);
        \coordinate (C) at (1,0);
        \coordinate (D) at (3/2,0);
        \draw[thick, black] (0.25,0) ellipse (0.45cm and 0.22cm);
        \draw[thick, black] (0.5,0) ellipse (0.8cm and 0.3cm);
        \roundrectangle{0}{0.75}{0}{2.2}{0.72}{0};
        \draw[thick] (A) circle (3.5pt);
        \draw[thick] (A) -- (B) -- (C) -- (D);
        \fill[black] (A) circle (2pt);
        \fill[black] (B) circle (2pt);
        \fill[black] (C) circle (2pt);
        \fill[black] (D) circle (2pt);
    \end{tikzpicture}};
\right) &= \tikz[baseline,yshift=5ex] \node[anchor=base] at (0,0) {\begin{tikzpicture}[scale=0.8]
        \coordinate (A) at (0,0);
        \coordinate (B) at (0,-1/2);
         \coordinate (C) at (0,-1);
         \coordinate (D) at (0,-3/2);
        \draw[thick] (A) -- (B) -- (C) -- (D);
        \node[fill=black, shape=rectangle, minimum size=1.5mm, inner sep=0mm] at (A) {};
        \fill[black] (B) circle (2pt);
        \fill[black] (C) circle (2pt);
        \fill[black] (D) circle (2pt);
    \end{tikzpicture}};,
    \end{align}
  where the top vertex of each poset corresponds to the root, which is represented as the rectangle around the graph $G$.
The set of tubings $\Gamma_G$ for a given graph $G$, together with inclusion, defines a partially ordered set which describes the boundary stratification of a convex polytope called the {\it graph associahedron of $G$} and denoted as $\mathcal{A}_G$. As such all boundaries of $\mathcal{A}_G$ are labelled by tubings, with the facets labelled by tubings containing a single tube, and the vertices labelled by maximal tubings. Two facets in $\mathcal{A}_G$ are adjacent, i.e.~intersect along a co-dimension-two boundary, if the tubes in their corresponding tubings are compatible. At the level of tubings the intersection of two compatible facets is given by taking the union of the tubings labelling them. The co-dimension of each boundary is specified by the cardinality of the corresponding tubing. In the following we will provide an explicit embedding of the graph associahedra that will realise the combinatorics described here. The graph associahedra contain familiar families of polytopes including:
\begin{itemize}
\item for the path graph with $n+1$ vertices, denoted by $P_{n+1}$, the graph associahedron of $P_{n+1}$ is the classical $n$-dimensional associahedron,
\item for the graph with $n+1$ vertices and $E_G=\emptyset$, denoted $N_{n+1}$, the graph associahedron is the $n$-dimensional simplex,
\item for the complete graph on $n+1$ vertices, denoted $K_{n+1}$, the graph associahedron is the $n$-dimensional permutohedron.
\end{itemize} 
An important property that we wish to emphasise is that all graph associahedra satisfy factorisation properties on their boundaries. Let the {\it reconnected complement} $G_t^*$ of a tube $t \in T_G$ be the graph with the vertex set $V_G \setminus t$ such that two vertices $a$ and $b$ are connected by an edge in $G_t^*$ if either of the induced graphs $G[\{a,b\}]$ or $G[\{a,b\} \cup t]$ is connected, as illustrated in the example below
\begin{equation}
G=\tikz[baseline,yshift=-0.90ex] \node[anchor=base] at (0,0) {\begin{tikzpicture}
        \coordinate (A) at (0,0);
        \coordinate (B) at (1/2,0);
        \coordinate (C) at (1/2,1/2);
        \coordinate (D) at (0,1/2);       
         \coordinate (E) at (1,1/2);
        \coordinate (F) at (1,0);
        \node at (-1/3,1/2){$t$};
        \draw[thick] (A) -- (B) -- (C) -- (D)--(A)--(C)--(E)--(F);
        \fill[black] (A) circle (2pt);
        \fill[black] (B) circle (2pt);
        \fill[black] (C) circle (2pt);
        \fill[black] (D) circle (2pt);
                \fill[black] (F) circle (2pt);
                \fill[black] (E) circle (2pt);
        \draw[thick, black] (0.25,1/2) ellipse (0.45cm and 0.22cm);
    \end{tikzpicture}}; \qquad \longrightarrow \qquad G_t^*=
    \tikz[baseline,yshift=-0.5ex] \node[anchor=base] at (0,0) {\begin{tikzpicture}
        \coordinate (A) at (0,0);
        \coordinate (B) at (1/2,0);
        \coordinate (C) at (1,1/2);
                \coordinate (D) at (1,0);
        \draw[thick] (A) -- (B) -- (C) -- (A);
                \draw[thick]  (C) -- (D);
        \fill[black] (A) circle (2pt);
        \fill[black] (B) circle (2pt);
        \fill[black] (C) circle (2pt);
                \fill[black] (D) circle (2pt);
    \end{tikzpicture}};.
\end{equation}
 Then the facet of the graph associahedron of $G$ corresponding to the tubing $\tau=\{t\}$, containing the single tube $t$, factorises as the Cartesian product of two simpler graph associahedra
\begin{align}
\partial_{t} (\mathcal{A}_G) = \mathcal{A}_{G[t]} \times \mathcal{A}_{G^*_t}.
\label{eq:frac_assoc}
\end{align}
Similar to the definition above, we can define the {\it reconnected complement for a tubing} $\tau=\{t_1,\ldots,t_{|\tau|}\}$ that contains pairwise non-adjacent and non-intersecting tubes, by iterating the above procedure for each $t \in \tau$. We denote the resulting graph as $G_{\tau}^*=((G^*_{t_1})^*_{t_2}\ldots)^*_{t_{|\tau|}}$.

\subsection{Amplitubes}
Having defined the graph associahedron of $G$ combinatorially we can proceed by associating the corresponding {\it amplitube} $A_G$, that is the function defined as 
 \begin{align}
 A_G = \sum_{\tau\in\Gamma_G^{\text{max}}} \prod_{t \in \tau} \frac{1}{X_t},
 \label{eq:amplitube}
 \end{align} 
where the sum is over all maximal tubings of $G$ and we have introduced a variable $X_t$ for each tube. We collectively refer to the set of all variables $\{X_t\}_{t\in T_G}$ as the {\it graph kinematic space} for graph $G$. 

As an example of an amplitube, let us consider the path graph $G=P_{n+1}$. If we label its vertices using labels $\{1,2,\ldots,n+1\}$ from left to right, then each tube is labelled by a string of consecutive integers $[i, j]=\{i,i+1,\ldots,j\}$ for $i<j$. Then, upon replacing $X_{[i, j ]} \rightarrow X_{i,j+2}$, where the latter are planar Mandelstam variables defined in \cite{Arkani-Hamed:2017mur}, the amplitube \eqref{eq:amplitube} recovers the familiar tree-level amplitudes in $\tr(\phi^3)$ theory. For example, the amplitube for the path graph on three vertices $P_3$ is given by 
\begin{align}
A_{P_3} = \frac{1}{X_{\{1\}}X_{\{1,2\}}}+\frac{1}{X_{\{2\}}X_{\{1,2\}}}+\frac{1}{X_{\{2\}}X_{\{2,3\}}}+\frac{1}{X_{\{3\}}X_{\{2,3\}}}+\frac{1}{X_{\{1\}}X_{\{3\}}}\,.
\end{align}
Another example is provided by the amplitube for the complete graph on three vertices $K_3$ that takes the following form
\begin{align}
A_{K_3} = \frac{1}{X_{\{1\}}X_{\{1,2\}}}+\frac{1}{X_{\{2\}}X_{\{1,2\}}}+\frac{1}{X_{\{2\}}X_{\{2,3\}}}+\frac{1}{X_{\{3\}}X_{\{2,3\}}}+\frac{1}{X_{\{1\}}X_{\{1,3\}}}+\frac{1}{X_{\{3\}}X_{\{1,3\}}}.
\end{align}
At the level of the amplitube, the factorisation property of the graph associahedra \eqref{eq:frac_assoc} is reflected by the following factorisation of its residue
\begin{align}
\underset{X_t=0}{\text{Res}} \left(A_G\right) = A_{G[t]} \cdot  A_{G^*_t}.
\end{align}
As an example, again considering the complete graph on three vertices $K_3$, we have
\begin{align}
\underset{X_{\{1,2\}}=0}{\text{Res}} \left( A_{K_3}\right) = \frac{1}{X_{\{1\}}}+ \frac{1}{X_{\{2\}}} = A_{P_2} \cdot A_{P_1},
\end{align}
which reflects the correct factorisation of the facet $X_{\{1,2\}}=0$ of the graph associahedron $\mathcal{A}_{K_3}$, since $K_3[\{1,2\}]=P_2$ and $(K_3)^*_{\{1,2\}}=P_1$.

To translate to the language of physics it is easy to see that the analogue of {\it locality} for the amplitubes is given by the fact that the variables appearing in the denominator of $A_G$ each correspond to {\it connected} subgraphs of $G$, whilst the analogue of {\it unitarity} is the fact that the residues of amplitubes are given by products of simpler amplitubes.

\subsection{Associahedron Embedding}\label{sec:ass_emb}
We will now turn the combinatorial statements of the last section into geometric ones by providing a realisation of the graph associahedra as convex polytopes. First, we want to point out that in the case of the path graph $P_{n+1}$, there exists an embedding intimately connected to the kinematics of $n$-particle scattering in $\tr(\phi^3)$ theory, commonly referred to as the ABHY associahedron. It is this construction which we wish to modify and then generalise to arbitrary graph associahedra. We will refer to the embedding introduced in this section as the {\it tube embedding}. Our prescription follows closely the one originally laid out for the associahedron in \cite{Arkani-Hamed:2017mur} together with earlier work on realisations of graph associahedra in \cite{devadoss2009realization}. Similar constructions have already appeared in the literature, for instance see \cite{He:2020onr,postnikov2009permutohedra}.

\begin{figure}[h]
    \centering
    \includegraphics[width=0.27\textwidth]{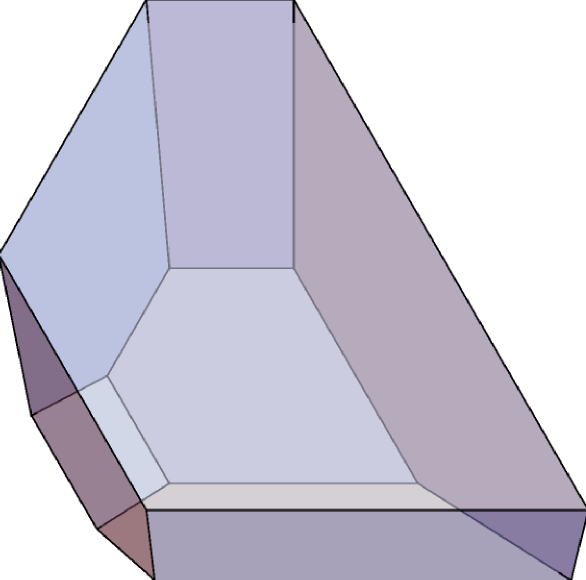} \quad \quad \includegraphics[width=0.31\textwidth]{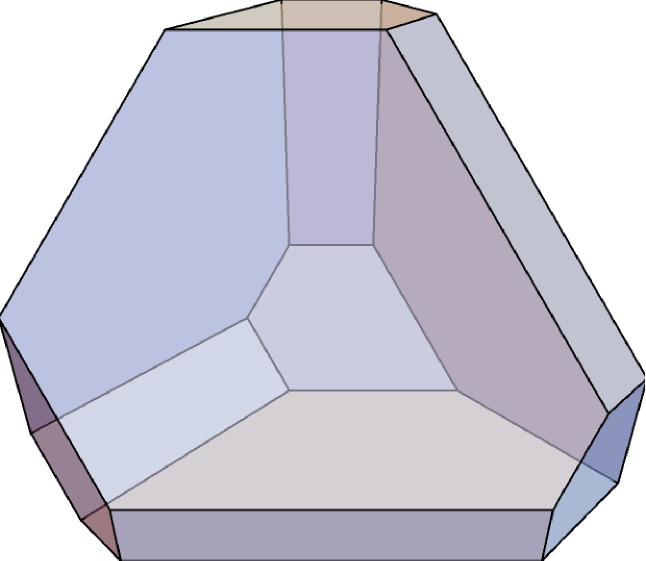} \quad \quad \includegraphics[width=0.3\textwidth]{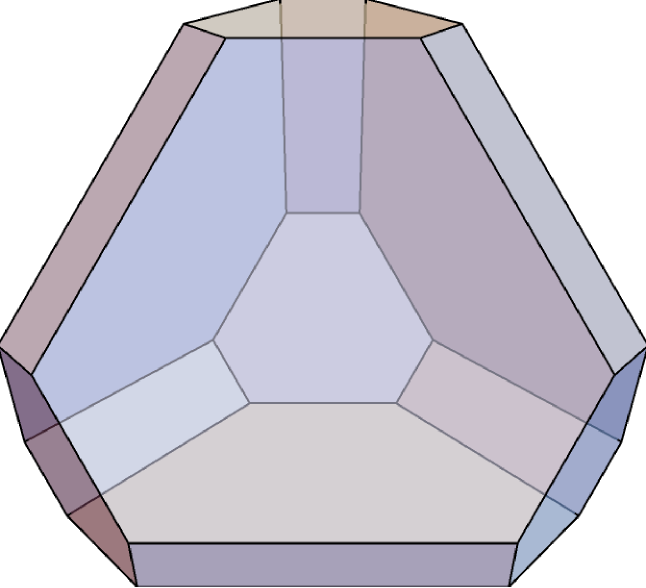}
    \caption{Embeddings of three-dimensional graph associahedra for the path graph, cycle graph, and complete graph on four vertices. These polytopes are examples of the three-dimensional associahedron, cyclohedron and permutohedron, respectively. }
    \label{fig:assoc_exam}
\end{figure}

We fix the graph $G$. For each tube $t \in T_G$, in addition to the kinematic variable $X_t$, we also introduce a {\it cut-parameter} $c_t$, which we take to be a positive real number. We impose that the kinematic variables satisfy the following linear constraints
\begin{align}
X_t = -\sum_{t' \subset t} c_t\,,
\label{eq:kin_constr}
\end{align}
where the sum runs over all tubes that are subsets of $t$. Additionally, we impose a similar relation also for the variable $X_{V_G}$ associated to the root:
\begin{equation}
X_{V_G}=-\sum_{t\in \overline{T}_G} c_t\,.
\end{equation} 
Importantly, in the case where the tube $t=\{v\}$ contains a single vertex $v$, we have $X_{\{ v \}} = -c_{\{ v \}}$, which allows one to re-express the linear relations \eqref{eq:kin_constr} as 
\begin{align}
X_t = \sum_{v \in t}X_{\{ v\} } -\sum_{t' \subset t, |t'|>1} c_t.
\end{align}
Here the sum over $c$'s runs over all tubes $t'$ that are subsets of $t$ and contain more than one vertex.  The tube embedding of the graph associahedron can now simply be stated in terms of the variables $X_{\{v\}}$  as
\begin{align}\label{eq_ABHY_assoc}
\boxed{\mathcal{A}_G=\{ (X_{\{1 \}},\ldots,X_{\{|V_G| \}}) \in \mathbb{R}^{|G|} :  (\forall_{ t \in T_G}  \, X_t \geq 0) \text{ and } (X_{V_G}=0)\}.}
\end{align}
In the case of the classical associahedra $\mathcal{A}_{P_{n+1}}$, corresponding to the path graph $G=P_{n+1}$, the tube embedding agrees with the one obtained by ABHY in \cite{Arkani-Hamed:2017mur}, after the latter is projected on an appropriate $n$ dimensional subspace\footnote{One needs to project the ABHY associahedron on the space parametrised by the planar Mandelstam variables $X_{i,i+2}$ for $i=1,2,\ldots,n-2$.} of the kinematic space. Examples of the tube embedding for the graph associahedra associated to the path graph $P_4$, the cycle graph $C_4$ and the complete graph $K_4$ are displayed in Fig.~\ref{fig:assoc_exam}.

To make a further comparison with the ABHY embedding, we can invert \eqref{eq:kin_constr} to obtain an expression for the $c$'s in terms of $X$'s. The expressions for the cut parameters then take the form of an alternating sum as 
\begin{align}
c_{t_0} =-X_{t_0} +\sum_{t_1 \subsetneq t_0}X_{t_1}-  \sum_{t_2 \subsetneq t_1 \subsetneq t_0}X_{t_2}+ \sum_{t_3 \subsetneq t_2 \subsetneq t_1 \subsetneq t_0}X_{t_3}- \ldots\,.
\end{align}
In the case of the classical $n$-dimensional associahedron $\mathcal{A}_{P_{n+1}}$, the above formula reduces to the familiar conditions
\begin{align}\label{eq:ABHY_const}
c_{[i,j]}=X_{[i-1,j]}+X_{[i,j-1]}-X_{[i,j]}-X_{[i-1,j-1]}\,,
\end{align}
where the $X_{[i,j]}$ can again be identified with the square of sums of consecutive momenta i.e. the planar Mandelstam variables.

We emphasize that the definition \eqref{eq_ABHY_assoc} is different from the one provided in the original ABHY construction for classical associahedra. In our case, the polytope $\mathcal{A}_G$ sits on a hyperplane $X_{V_G}=0$ inside a $|V_G|$-dimensional Euclidean space. In the original construction, the space was parametrised by $\binom{n}{2}$ planar Mandelstam variables, and the polytope was residing on the intersection of the conditions \eqref{eq:ABHY_const}. Our construction also reduces to the one provided in \cite{postnikov2009permutohedra} when all cut-parameters $c_t$ are set to 1.

Finally, every graph associahedron $\mathcal{A}_G$ is a simple convex polytope, i.e.~a $d$-dimensional polytope with all vertices incident to exactly $d$ facets. Therefore one can find its canonical differential form using the general formula
\begin{align}
\Omega\left( \mathcal{A}_G \right)= \sum_{\tau\in \Gamma_G^{\text{max}}} \text{sign}_\tau \bigwedge_{t \in \tau} \dd\log(X_t),
\label{eq:can_form_assoc}
\end{align}
where the sum runs over all vertices of $\mathcal{A}_G$, and the signs can be determined by requiring that the canonical form $\Omega\left( \mathcal{A}_G \right)$ is projectively invariant. The amplitube \eqref{eq:amplitube} can then be recovered from the canonical forms as
\begin{equation}
\Omega(\mathcal{A}_G)=A_G \, \dd\mu_G\,,
\end{equation}
where $\dd \mu_G=\dd X_{\{1\}}\wedge \dd X_{\{2\}} \wedge \ldots\wedge \dd X_{\{|V_G|-1\}}$ and we solved the condition $X_{V_G}=0$ to eliminate $X_{\{|V_G|\}}$ from our expressions.

\subsection{Examples of Graph Associahedra}\label{sec:ass_exam}
Before moving on to discuss graph cosmohedra, we take a closer look at some two-dimensional examples of graph associahedra that correspond to graphs with three vertices, which we will return to in the next section. Consider first the empty graph $N_3$ on three vertices, i.e. $V_{N_3}=\{1,2,3\}$ and $E_{N_3}=\emptyset$. The inequalities defining the graph associahedron $\mathcal{A}_{N_3}$ in this case are
\begin{align}
X_{\{1\}} \geq 0, \quad X_{\{2\}} \geq 0, \quad X_{\{3\}} \geq 0, \quad X_{\{1\}}+X_{\{2\}}+X_{\{3\}}=c_{\{1,2,3 \} },
\label{eq:v3_empty}
\end{align}
and the resulting polytope is a two-dimensional simplex, i.e.~a triangle. Next, consider the path graph $P_3$ on three vertices. In this case the inequalities that define the graph associahedron $\mathcal{A}_{P_3}$ are given by 
\begin{align}
&X_{\{1\}} \geq 0, \quad X_{\{2\}} \geq 0, \quad X_{\{3\}} \geq 0, \notag \\
&X_{\{1\}}+X_{\{2\}} \geq c_{\{1,2\}}, \quad  X_{\{2\}}+X_{\{3\}} \geq c_{\{2,3\}}, \notag \\
& X_{\{1\}}+X_{\{2\}}+X_{\{3\}} = c_{\{1,2\}}+c_{\{2,3\}}+c_{\{1,2,3\}}.
\label{eq:v3_path}
\end{align}
These inequalities carve out a pentagon i.e.~the classical two-dimensional associahedron. Finally, consider the complete graph $K_3$ on three vertices. The resulting set of inequalities is
\begin{align}
&X_{\{1\}} \geq 0, \quad X_{\{2\}} \geq 0, \quad X_{\{3\}} \geq 0, \notag \\
&X_{\{1\}}+X_{\{2\}} \geq c_{\{1,2\}}, \quad X_{\{2\}}+X_{\{3\}} \geq c_{\{2,3\}}, \quad X_{\{1\}}+X_{\{3\}} \geq c_{\{1,3\}}, \notag \\
& X_{\{1\}}+X_{\{2\}}+X_{\{3\}} = c_{\{1,2\}}+c_{\{2,3\}}+c_{\{1,3\}}+c_{\{1,2,3\}},
\end{align}
which produce a hexagon i.e.~the two-dimensional permutohedron. We provide a list of combinatorial data associated to graph associahedra for various graphs with two, three and four vertices in Table \ref{tab:assoc}.

\begin{table}
\begin{center}
\begin{tabular}{c|c|c|c|c}
Graph $G$&dimension& codim-1&codim-2&codim-3\\
\hline
$\tikz[baseline,yshift=0.5ex] \node[anchor=base] at (0,0) {\begin{tikzpicture}
        \coordinate (A) at (0,0);
        \coordinate (B) at (1/2,0);
        \fill[black] (A) circle (2pt);
        \fill[black] (B) circle (2pt);
    \end{tikzpicture}};$ &1&2&&\\
\hline
$\tikz[baseline,yshift=0.5ex] \node[anchor=base] at (0,0) {\begin{tikzpicture}
        \coordinate (A) at (0,0);
        \coordinate (B) at (1/2,0);
        \draw[thick] (A) -- (B);
        \fill[black] (A) circle (2pt);
        \fill[black] (B) circle (2pt);
    \end{tikzpicture}};$ &1&2&&\\
\hline
\hline
$\tikz[baseline,yshift=0.5ex] \node[anchor=base] at (0,0) {\begin{tikzpicture}
        \coordinate (A) at (0,0);
        \coordinate (B) at (1/2,0);
        \coordinate (C) at (1,0);
        \fill[black] (A) circle (2pt);
        \fill[black] (B) circle (2pt);
        \fill[black] (C) circle (2pt);
    \end{tikzpicture}};$ &2&3&3&\\
    \hline
$\tikz[baseline,yshift=0.5ex] \node[anchor=base] at (0,0) {\begin{tikzpicture}
        \coordinate (A) at (0,0);
        \coordinate (B) at (1/2,0);
        \coordinate (C) at (1,0);
        \draw[thick] (A) -- (B) ;
        \fill[black] (A) circle (2pt);
        \fill[black] (B) circle (2pt);
        \fill[black] (C) circle (2pt);
    \end{tikzpicture}};$ &2&4&4&\\
    \hline
$\tikz[baseline,yshift=0.5ex] \node[anchor=base] at (0,0) {\begin{tikzpicture}
        \coordinate (A) at (0,0);
        \coordinate (B) at (1/2,0);
        \coordinate (C) at (1,0);
        \draw[thick] (A) -- (B) -- (C);
        \fill[black] (A) circle (2pt);
        \fill[black] (B) circle (2pt);
        \fill[black] (C) circle (2pt);
    \end{tikzpicture}};$ &2&5&5&\\
        \hline
$\tikz[baseline,yshift=0.5ex] \node[anchor=base] at (0,0) {\begin{tikzpicture}
        \coordinate (A) at (0,0);
        \coordinate (B) at (1/2,0);
        \coordinate (C) at (1/4,1/3);
        \draw[thick] (A) -- (B) -- (C) -- (A);
        \fill[black] (A) circle (2pt);
        \fill[black] (B) circle (2pt);
        \fill[black] (C) circle (2pt);
    \end{tikzpicture}};$ &2&6&6&\\
    \hline
    \hline
$\tikz[baseline,yshift=0.5ex] \node[anchor=base] at (0,0) {\begin{tikzpicture}
        \coordinate (A) at (0,0);
        \coordinate (B) at (1/2,0);
        \coordinate (C) at (1,0);
        \coordinate (D) at (3/2,0);
        \draw[thick] (A) -- (B) -- (C) -- (D);
        \fill[black] (A) circle (2pt);
        \fill[black] (B) circle (2pt);
        \fill[black] (C) circle (2pt);
        \fill[black] (D) circle (2pt);
    \end{tikzpicture}};$ &3&9&21&14\\
\hline
$\tikz[baseline,yshift=0.5ex] \node[anchor=base] at (0,0) {\begin{tikzpicture}
       \coordinate (A) at (0,0);
        \coordinate (B) at (1/4,1/4);
        \coordinate (C) at (-1/4,1/4);
        \coordinate (D) at (0,-1/3);
        \draw[thick] (B) -- (A) -- (C);
                \draw[thick] (D) -- (A);
        \fill[black] (A) circle (2pt);
        \fill[black] (B) circle (2pt);
        \fill[black] (C) circle (2pt);
        \fill[black] (D) circle (2pt);
    \end{tikzpicture}};$ &3&10&24&16\\
\hline
$\tikz[baseline,yshift=0.5ex] \node[anchor=base] at (0,0) {\begin{tikzpicture}
        \coordinate (A) at (0,0);
        \coordinate (B) at (1/2,0);
        \coordinate (C) at (1/2,1/2);
        \coordinate (D) at (0,1/2);
        \draw[thick] (A) -- (B) -- (C) -- (D)--(A);
        \fill[black] (A) circle (2pt);
        \fill[black] (B) circle (2pt);
        \fill[black] (C) circle (2pt);
        \fill[black] (D) circle (2pt);
    \end{tikzpicture}};$ &3&12&30&20\\
    \hline
$\tikz[baseline,yshift=0.5ex] \node[anchor=base] at (0,0) {\begin{tikzpicture}
        \coordinate (A) at (0,0);
        \coordinate (B) at (1/2,0);
        \coordinate (C) at (1/2,1/2);
        \coordinate (D) at (0,1/2);
        \draw[thick] (A) -- (B) -- (C) -- (D)--(A);
        \draw[thick] (A) -- (C);
        \draw[thick]  (B) --(D);
        \fill[black] (A) circle (2pt);
        \fill[black] (B) circle (2pt);
        \fill[black] (C) circle (2pt);
        \fill[black] (D) circle (2pt);
    \end{tikzpicture}};$ &3&14&36&24\\
\end{tabular}
\end{center}
\caption{The number of boundaries of various co-dimension of the associahedron on graph $G$.}
\label{tab:assoc}
\end{table}

\section{Graph Cosmohedra}\label{sec:graph_cosmo}
We now wish to promote the amplitubes introduced in the last section into a cosmological setting, the result of which we will refer to as {\it cosmological amplitubes}. As we shall explain these functions are most naturally written in terms of {\it regions} and {\it regional tubings} of the graph, which we introduce. In the case of the path graph the expression for the cosmological amplitube reduces to that of the wavefunction for $\tr (\phi^3)$ theory. As was the case for the graph associahedra and the amplitubes in the previous section, we will see that the cosmological amplitubes also have a geometric counterpart we refer to as {\it graph cosmohedra}. Graph cosmohedra are a new class of convex polytope that interpolate between the permutohedron and the permutoassociahedron. We will provide explicit embeddings for the graph cosmohedra which will rely on the already introduced variables for the graph associahedra. In the case of the path graph, the graph cosmohedron and its embedding recovers the recently introduced cosmohedron and the embedding provided in \cite{Arkani-Hamed:2024jbp}.

\subsection{Regions and Regional Tubings}
In the case of the path graph our goal is to reproduce the {\it cosmohedron} of \cite{Arkani-Hamed:2024jbp} whose combinatorics is governed by sub-polygons and their nested polyangulations. Therefore, in order to extend this to an arbitrary graph we will need to generalise the notions of sub-polygons and nested polyangulations to {\it regions} and {\it nested regionalisations}. These will become the {\it regional tubes} and {\it regional tubings} we now introduce.

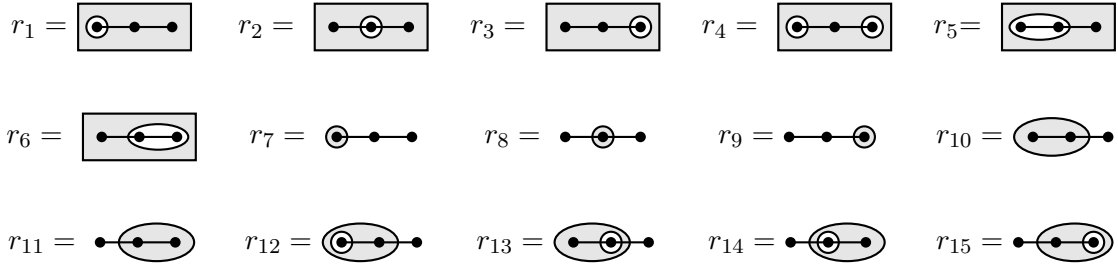
\begin{figure}[t!]
    \centering
    \begin{tabular}{c c c c c} 
        $r_1=$\tikz[baseline,yshift=0.5ex] \node[anchor=base] at (0,0) {\begin{tikzpicture}
        \coordinate (A) at (0,0);
        \coordinate (B) at (1/2,0);
        \coordinate (C) at (1,0);
        \coordinate (D) at (3/2,0);
        \fillroundrectangle{0}{0.5}{0}{1.5}{0.62}{0};
        \roundrectangle{0}{0.5}{0}{1.5}{0.62}{0};
        \filldraw[thick,white] (A) circle (4pt);
        \draw[thick] (A) circle (4pt);
        \draw[thick] (A) -- (B) -- (C);
        \fill[black] (A) circle (2pt);
        \fill[black] (B) circle (2pt);
        \fill[black] (C) circle (2pt);
    \end{tikzpicture}}; &
      $r_2=$  \tikz[baseline,yshift=0.5ex] \node[anchor=base] at (0,0) {\begin{tikzpicture}
        \coordinate (A) at (0,0);
        \coordinate (B) at (1/2,0);
        \coordinate (C) at (1,0);
        \coordinate (D) at (3/2,0);
        \fillroundrectangle{0}{0.5}{0}{1.5}{0.62}{0};
        \roundrectangle{0}{0.5}{0}{1.5}{0.62}{0};
        \filldraw[thick,white] (B) circle (4pt);
        \draw[thick] (B) circle (4pt);
        \draw[thick] (A) -- (B) -- (C);
        \fill[black] (A) circle (2pt);
        \fill[black] (B) circle (2pt);
        \fill[black] (C) circle (2pt);
    \end{tikzpicture}}; &
       $r_3=$ \tikz[baseline,yshift=0.5ex] \node[anchor=base] at (0,0) {\begin{tikzpicture}
        \coordinate (A) at (0,0);
        \coordinate (B) at (1/2,0);
        \coordinate (C) at (1,0);
        \coordinate (D) at (3/2,0);
        \fillroundrectangle{0}{0.5}{0}{1.5}{0.62}{0};
        \roundrectangle{0}{0.5}{0}{1.5}{0.62}{0};
        \filldraw[thick,white] (C) circle (4pt);
        \draw[thick] (C) circle (4pt);
        \draw[thick] (A) -- (B) -- (C);
        \fill[black] (A) circle (2pt);
        \fill[black] (B) circle (2pt);
        \fill[black] (C) circle (2pt);
    \end{tikzpicture}}; &
       $r_4=$  \tikz[baseline,yshift=0.5ex] \node[anchor=base] at (0,0) {\begin{tikzpicture}
        \coordinate (A) at (0,0);
        \coordinate (B) at (1/2,0);
        \coordinate (C) at (1,0);
        \coordinate (D) at (3/2,0);
        \fillroundrectangle{0}{0.5}{0}{1.5}{0.62}{0};
        \roundrectangle{0}{0.5}{0}{1.5}{0.62}{0};
        \filldraw[thick,white] (A) circle (4pt);
        \filldraw[thick,white] (C) circle (4pt);
        \draw[thick] (A) circle (4pt);
        \draw[thick] (C) circle (4pt);
        \draw[thick] (A) -- (B) -- (C);
        \fill[black] (A) circle (2pt);
        \fill[black] (B) circle (2pt);
        \fill[black] (C) circle (2pt);
    \end{tikzpicture}};  &
      $r_5$=  \tikz[baseline,yshift=0.5ex] \node[anchor=base] at (0,0) {\begin{tikzpicture}
        \coordinate (A) at (0,0);
        \coordinate (B) at (1/2,0);
        \coordinate (C) at (1,0);
        \coordinate (D) at (3/2,0);
        \fillroundrectangle{0}{0.5}{0}{1.5}{0.62}{0};
        \roundrectangle{0}{0.5}{0}{1.5}{0.62}{0};
        \filldraw[thick, white] (0.25,0) ellipse (0.4cm and 0.17cm);
        \draw[thick,black] (0.25,0) ellipse (0.4cm and 0.17cm);
        \draw[thick] (A) -- (B) -- (C);
        \fill[black] (A) circle (2pt);
        \fill[black] (B) circle (2pt);
        \fill[black] (C) circle (2pt);
    \end{tikzpicture}}; \\[2em]
       $r_6=$ \tikz[baseline,yshift=0.5ex] \node[anchor=base] at (0,0) {\begin{tikzpicture}
        \coordinate (A) at (0,0);
        \coordinate (B) at (1/2,0);
        \coordinate (C) at (1,0);
        \coordinate (D) at (3/2,0);
        \fillroundrectangle{0}{0.5}{0}{1.5}{0.62}{0};
        \roundrectangle{0}{0.5}{0}{1.5}{0.62}{0};
        \filldraw[thick, white] (0.75,0) ellipse (0.4cm and 0.17cm);
        \draw[thick,black] (0.75,0) ellipse (0.4cm and 0.17cm);
        \draw[thick] (A) -- (B) -- (C);
        \fill[black] (A) circle (2pt);
        \fill[black] (B) circle (2pt);
        \fill[black] (C) circle (2pt);
    \end{tikzpicture}}; &
  $r_7=$  \tikz[baseline,yshift=0.5ex] \node[anchor=base] at (0,0) {\begin{tikzpicture}
        \coordinate (A) at (0,0);
        \coordinate (B) at (1/2,0);
        \coordinate (C) at (1,0);
        \coordinate (D) at (3/2,0);
        \fill[gray!20] (A) circle (4pt);
        \draw[black,thick] (A) circle (4pt);
        \draw[thick] (A) -- (B) -- (C);
        \fill[black] (A) circle (2pt);
        \fill[black] (B) circle (2pt);
        \fill[black] (C) circle (2pt);
    \end{tikzpicture}}; &
   $r_8=$ \tikz[baseline,yshift=0.5ex] \node[anchor=base] at (0,0) {\begin{tikzpicture}
        \coordinate (A) at (0,0);
        \coordinate (B) at (1/2,0);
        \coordinate (C) at (1,0);
        \coordinate (D) at (3/2,0);
        \fill[gray!20] (B) circle (4pt);
        \draw[black,thick] (B) circle (4pt);
        \draw[thick] (A) -- (B) -- (C);
        \fill[black] (A) circle (2pt);
        \fill[black] (B) circle (2pt);
        \fill[black] (C) circle (2pt);
    \end{tikzpicture}}; &
    $r_9=$\tikz[baseline,yshift=0.5ex] \node[anchor=base] at (0,0) {\begin{tikzpicture}
        \coordinate (A) at (0,0);
        \coordinate (B) at (1/2,0);
        \coordinate (C) at (1,0);
        \coordinate (D) at (3/2,0);
        \fill[gray!20] (C) circle (4pt);
        \draw[black,thick] (C) circle (4pt);
        \draw[thick] (A) -- (B) -- (C);
        \fill[black] (A) circle (2pt);
        \fill[black] (B) circle (2pt);
        \fill[black] (C) circle (2pt);
    \end{tikzpicture}}; &
    $r_{10}=$\tikz[baseline,yshift=0.5ex] \node[anchor=base] at (0,0) {\begin{tikzpicture}
        \coordinate (A) at (0,0);
        \coordinate (B) at (1/2,0);
        \coordinate (C) at (1,0);
        \coordinate (D) at (3/2,0);
        \filldraw[thick, gray!20] (0.25,0) ellipse (0.5cm and 0.25cm);
        \draw[thick,black] (0.25,0) ellipse (0.5cm and 0.25cm);
        \draw[thick] (A) -- (B) -- (C);
        \fill[black] (A) circle (2pt);
        \fill[black] (B) circle (2pt);
        \fill[black] (C) circle (2pt);
    \end{tikzpicture}}; \\[2em]
    $r_{11}=$ \tikz[baseline,yshift=0.5ex] \node[anchor=base] at (0,0) {\begin{tikzpicture}
        \coordinate (A) at (0,0);
        \coordinate (B) at (1/2,0);
        \coordinate (C) at (1,0);
        \coordinate (D) at (3/2,0);
        \filldraw[thick, gray!20] (0.75,0) ellipse (0.5cm and 0.25cm);
        \draw[thick,black] (0.75,0) ellipse (0.5cm and 0.25cm);
        \draw[thick] (A) -- (B) -- (C);
        \fill[black] (A) circle (2pt);
        \fill[black] (B) circle (2pt);
        \fill[black] (C) circle (2pt);
    \end{tikzpicture}}; &
    $r_{12}=$\tikz[baseline,yshift=0.5ex] \node[anchor=base] at (0,0) {\begin{tikzpicture}
        \coordinate (A) at (0,0);
        \coordinate (B) at (1/2,0);
        \coordinate (C) at (1,0);
        \coordinate (D) at (3/2,0);
        \filldraw[thick, gray!20] (0.25,0) ellipse (0.5cm and 0.25cm);
        \fill[white] (A) circle (4pt);
        \draw[black,thick] (A) circle (4pt);
        \draw[thick,black] (0.25,0) ellipse (0.5cm and 0.25cm);
        \draw[thick] (A) -- (B) -- (C);
        \fill[black] (A) circle (2pt);
        \fill[black] (B) circle (2pt);
        \fill[black] (C) circle (2pt);
    \end{tikzpicture}}; &
    $r_{13}=$\tikz[baseline,yshift=0.5ex] \node[anchor=base] at (0,0) {\begin{tikzpicture}
        \coordinate (A) at (0,0);
        \coordinate (B) at (1/2,0);
        \coordinate (C) at (1,0);
        \coordinate (D) at (3/2,0);
        \filldraw[thick, gray!20] (0.25,0) ellipse (0.5cm and 0.25cm);
        \fill[white] (B) circle (4pt);
        \draw[black,thick] (B) circle (4pt);
        \draw[thick,black] (0.25,0) ellipse (0.5cm and 0.25cm);
        \draw[thick] (A) -- (B) -- (C);
        \fill[black] (A) circle (2pt);
        \fill[black] (B) circle (2pt);
        \fill[black] (C) circle (2pt);
    \end{tikzpicture}}; &
    $r_{14}=$\tikz[baseline,yshift=0.5ex] \node[anchor=base] at (0,0) {\begin{tikzpicture}
        \coordinate (A) at (0,0);
        \coordinate (B) at (1/2,0);
        \coordinate (C) at (1,0);
        \coordinate (D) at (3/2,0);
        \filldraw[thick, gray!20] (0.75,0) ellipse (0.5cm and 0.25cm);
        \fill[white] (B) circle (4pt);
        \draw[black,thick] (B) circle (4pt);
        \draw[thick,black] (0.75,0) ellipse (0.5cm and 0.25cm);
        \draw[thick] (A) -- (B) -- (C);
        \fill[black] (A) circle (2pt);
        \fill[black] (B) circle (2pt);
        \fill[black] (C) circle (2pt);
    \end{tikzpicture}}; &
    $r_{15}=$\tikz[baseline,yshift=0.5ex] \node[anchor=base] at (0,0) {\begin{tikzpicture}
        \coordinate (A) at (0,0);
        \coordinate (B) at (1/2,0);
        \coordinate (C) at (1,0);
        \coordinate (D) at (3/2,0);
        \filldraw[thick, gray!20] (0.75,0) ellipse (0.5cm and 0.25cm);
        \fill[white] (C) circle (4pt);
        \draw[black,thick] (C) circle (4pt);
        \draw[thick,black] (0.75,0) ellipse (0.5cm and 0.25cm);
        \draw[thick] (A) -- (B) -- (C);
        \fill[black] (A) circle (2pt);
        \fill[black] (B) circle (2pt);
        \fill[black] (C) circle (2pt);
    \end{tikzpicture}}; 
    \end{tabular}
    \caption{The fifteen regions for the graph $P_3$.}
    \label{fig:regions_P3}
\end{figure}

We say that $r=\{t_1,t_2,\ldots,t_{|r|}\}\subset \overline{T}_G$ is a {\it region} of the graph $G$ if 
\begin{itemize}
\item $r$ is a subset of $\overline{T}_G$ such that all tubes of $r$ are compatible as tubes,
\item the poset defined by inclusion on the elements of $r$ has a single maximal element,
\item all other tubes in $r$ are compatible with each other and not nested. 
\end{itemize} We refer to the element in $r$ that is maximal with respect to inclusion as the parent $t_p^{(r)}$ and all other elements as the children $t_{c,i}^{(r)}$. Note, the parent will either correspond to a tube $t_p^{(r)} \in T_G$ or to the root $t_p^{(r)}=V_G$. In other words every region $r$ is a graded poset with at most two layers, with one parent and a (possibly empty) set of children.  Every tubing $\tau\in \Gamma_G$ defines a collection of regions that we denote as $R(\tau)$. To construct the set $R(\tau)$ we take for each $t\in \overline{\tau}$ the region with $t$ as the parent together with all its children $t \supsetneq t' \in \tau$. An example of this procedure is given by the following
\begin{equation}
R \left(   \tikz[baseline,yshift=0.5ex] \node[anchor=base] at (0,0) {\begin{tikzpicture}
        \coordinate (A) at (0,0);
        \coordinate (B) at (1/2,0);
        \coordinate (C) at (1,0);
        \coordinate (D) at (3/2,0);
                \roundrectangle{0}{0.5}{0}{1.6}{0.62}{0};

        \fill[white] (A) circle (4pt);
        \draw[black,thick] (A) circle (4pt);
        \draw[thick,black] (0.25,0) ellipse (0.5cm and 0.25cm);
        \draw[thick] (A) -- (B) -- (C);
        \fill[black] (A) circle (2pt);
        \fill[black] (B) circle (2pt);
        \fill[black] (C) circle (2pt);
    \end{tikzpicture}}; \right) = \left\{ \tikz[baseline,yshift=0.5ex] \node[anchor=base] at (0,0) {\begin{tikzpicture}
        \coordinate (A) at (0,0);
        \coordinate (B) at (1/2,0);
        \coordinate (C) at (1,0);
        \coordinate (D) at (3/2,0);
         \fillroundrectangle{0}{0.5}{0}{1.6}{0.62}{0};
          \roundrectangle{0}{0.5}{0}{1.6}{0.62}{0};
        \filldraw[thick, white] (0.25,0) ellipse (0.5cm and 0.25cm);
        \draw[thick,black] (0.25,0) ellipse (0.5cm and 0.25cm);        
        \draw[thick] (A) -- (B) -- (C);
        \fill[black] (A) circle (2pt);
        \fill[black] (B) circle (2pt);
        \fill[black] (C) circle (2pt);
    \end{tikzpicture}};,\tikz[baseline,yshift=0.5ex] \node[anchor=base] at (0,0) {\begin{tikzpicture}
        \coordinate (A) at (0,0);
        \coordinate (B) at (1/2,0);
        \coordinate (C) at (1,0);
        \coordinate (D) at (3/2,0);
        \filldraw[thick, gray!20] (0.25,0) ellipse (0.5cm and 0.25cm);
        \draw[thick,black] (0.25,0) ellipse (0.5cm and 0.25cm);   
        \fill[white] (A) circle (4pt);
        \draw[black,thick] (A) circle (4pt);     
        \draw[thick] (A) -- (B) -- (C);
        \fill[black] (A) circle (2pt);
        \fill[black] (B) circle (2pt);
        \fill[black] (C) circle (2pt);
    \end{tikzpicture}};,\tikz[baseline,yshift=0.5ex] \node[anchor=base] at (0,0) {\begin{tikzpicture}
        \coordinate (A) at (0,0);
        \coordinate (B) at (1/2,0);
        \coordinate (C) at (1,0);
        \coordinate (D) at (3/2,0);
        \fill[gray!20] (A) circle (4pt);
        \draw[black,thick] (A) circle (4pt);     
        \draw[thick] (A) -- (B) -- (C);
        \fill[black] (A) circle (2pt);
        \fill[black] (B) circle (2pt);
        \fill[black] (C) circle (2pt);
    \end{tikzpicture}}; \right\}.
\end{equation}
Although the notion of the regions is well defined for any graph, we can get an intuitive understanding by considering planar graphs. In this case, we can depict the regions by shading in the area between the parent and the children in the planar embedding of the graph. For example, for the path graph $P_3$ there are $15$ possible regions, six of which have the root as the parent, as depicted in Fig.~\ref{fig:regions_P3}. Moreover, since the graph $P_3$ is related to the two-dimensional associahedron that labels triangulations of a pentagon, there is a natural bijection from the set of regions of $P_3$ to the set of subpolygons of a pentagon.

\begin{figure}
\center
\includegraphics[scale=1]{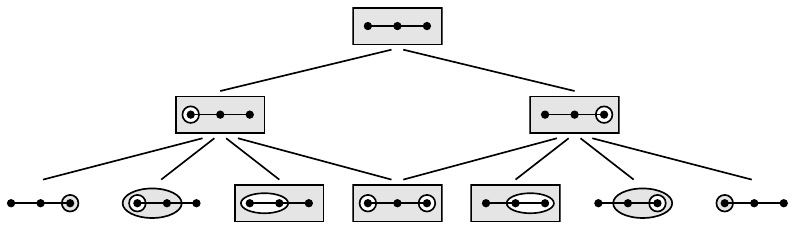}
\caption{An example of the sub-region relation $\prec_{\text{reg}}$ for a subset of the regions of $P_3$.}
\label{fig:reg_rel}
\end{figure}

We say two regions $r$ and $r'$ of $G$ are {\it compatible} if we have 
\begin{align}
(r \cup r')\setminus V_G \in \Gamma_G,
\end{align}
which means that all tubes in the union of $r$ and $r'$ form a tubing of $G$. A region $r'$ is a {\it sub-region} of $r$ if it can be realised as a region on the graph  $G^{(r)}:=G[t_p^{(r)}]^*_{\{t_{c,i}^{(r)}\}}$, that is a region in the reconnected complement in the graph induced by the parent tube $t_p^{(r)}$ of all children tubes $t_{c,i}^{(r)}$. An example of the sub-region relation, denoted $r_1 \prec_{\text{reg}} r_2$ for $r_1$ is a sub-region of $r_2$, on a subset of the regions of $P_3$ is given in Fig.~\ref{fig:reg_rel}.

Given two compatible regions that are not sub-regions of one another, we refer to their intersection as the {\it border}. The border will necessary be empty or contain a single tube. The set of vertices $V^{(r)}=t_p^{(r)}\setminus \bigcup_{i} t_{c,i}^{(r)}$ {\it covered} by a region $r$ is defined as the complement of the set of vertices in the children inside the set of vertices of the parent. A collection of sub-regions $\{r_i\}$ of $r$ is said to {\it cover} $r$ if $V^{(r)}=\cup_i V^{(r_i)}$ and $V^{(r_i)}$ are mutually disjoint. When we have such a covering, we use the notation $r = \sqcup_i r_i$. Again, these definition are much easier to visualise in a simple case, for example
\begin{align}\label{eq:subregions_1}
\tikz[baseline,yshift=0.5ex] \node[anchor=base] at (0,0) {\begin{tikzpicture}
        \coordinate (A) at (0,0);
        \coordinate (B) at (1/2,0);
        \coordinate (C) at (1,0);
        \coordinate (D) at (3/2,0);
        \fillroundrectangle{0}{0.8}{0}{2.2}{0.62}{0};
         \roundrectangle{0}{0.8}{0}{2.2}{0.62}{0};
        \filldraw[thick, white] (0.75,0) ellipse (0.5cm and 0.15cm);
        \draw[thick, black] (0.75,0) ellipse (0.5cm and 0.15cm);
         \draw[thick] (A) -- (B) -- (C) -- (D);
        \fill[black] (A) circle (2pt);
        \fill[black] (B) circle (2pt);
        \fill[black] (C) circle (2pt);
        \fill[black] (D) circle (2pt);
    \end{tikzpicture}}; = \tikz[baseline,yshift=0.5ex] \node[anchor=base] at (0,0) {\begin{tikzpicture}
        \coordinate (A) at (0,0);
        \coordinate (B) at (1/2,0);
        \coordinate (C) at (1,0);
        \coordinate (D) at (3/2,0);
        \fillroundrectangle{0}{0.8}{0}{2.2}{0.62}{0};
         \roundrectangle{0}{0.8}{0}{2.2}{0.62}{0};
        \filldraw[thick, white] (0.5,0) ellipse (0.75cm and 0.15cm);
        \draw[thick, black] (0.5,0) ellipse (0.75cm and 0.2cm);
         \draw[thick] (A) -- (B) -- (C) -- (D);
        \fill[black] (A) circle (2pt);
        \fill[black] (B) circle (2pt);
        \fill[black] (C) circle (2pt);
        \fill[black] (D) circle (2pt);
    \end{tikzpicture}}; \sqcup  
    \tikz[baseline,yshift=0.5ex] \node[anchor=base] at (0,0) {\begin{tikzpicture}
        \coordinate (A) at (0,0);
        \coordinate (B) at (1/2,0);
        \coordinate (C) at (1,0);
        \coordinate (D) at (3/2,0);
        \filldraw[thick, gray!20] (0.5,0) ellipse (0.8cm and 0.25cm);
        \draw[thick, black] (0.5,0) ellipse (0.8cm and 0.25cm);
        \filldraw[thick, white] (0.75,0) ellipse (0.45cm and 0.15cm);
        \draw[thick, black] (0.75,0) ellipse (0.45cm and 0.15cm);
         \draw[thick] (A) -- (B) -- (C) -- (D);
        \fill[black] (A) circle (2pt);
        \fill[black] (B) circle (2pt);
        \fill[black] (C) circle (2pt);
        \fill[black] (D) circle (2pt);
    \end{tikzpicture}};,
    \end{align}
 An alternative example of a covering of the same region is given by\footnote{Relations \eqref{eq:subregions_1} and \eqref{eq:subregions_2} can be interpreted as two triangulations of a particular quadrilateral inside a hexagon.}
 \begin{align}\label{eq:subregions_2}
\tikz[baseline,yshift=0.5ex] \node[anchor=base] at (0,0) {\begin{tikzpicture}
        \coordinate (A) at (0,0);
        \coordinate (B) at (1/2,0);
        \coordinate (C) at (1,0);
        \coordinate (D) at (3/2,0);
        \fillroundrectangle{0}{0.8}{0}{2.2}{0.62}{0};
         \roundrectangle{0}{0.8}{0}{2.2}{0.62}{0};
        \filldraw[thick, white] (0.75,0) ellipse (0.5cm and 0.15cm);
        \draw[thick, black] (0.75,0) ellipse (0.5cm and 0.15cm);
         \draw[thick] (A) -- (B) -- (C) -- (D);
        \fill[black] (A) circle (2pt);
        \fill[black] (B) circle (2pt);
        \fill[black] (C) circle (2pt);
        \fill[black] (D) circle (2pt);
    \end{tikzpicture}}; = \tikz[baseline,yshift=0.5ex] \node[anchor=base] at (0,0) {\begin{tikzpicture}
        \coordinate (A) at (0,0);
        \coordinate (B) at (1/2,0);
        \coordinate (C) at (1,0);
        \coordinate (D) at (3/2,0);
        \fillroundrectangle{0}{0.8}{0}{2.2}{0.62}{0};
         \roundrectangle{0}{0.8}{0}{2.2}{0.62}{0};
        \filldraw[thick, white] (1,0) ellipse (0.75cm and 0.15cm);
        \draw[thick, black] (1,0) ellipse (0.75cm and 0.2cm);
         \draw[thick] (A) -- (B) -- (C) -- (D);
        \fill[black] (A) circle (2pt);
        \fill[black] (B) circle (2pt);
        \fill[black] (C) circle (2pt);
        \fill[black] (D) circle (2pt);
    \end{tikzpicture}};  \sqcup \tikz[baseline,yshift=0.5ex] \node[anchor=base] at (0,0) {\begin{tikzpicture}
        \coordinate (A) at (0,0);
        \coordinate (B) at (1/2,0);
        \coordinate (C) at (1,0);
        \coordinate (D) at (3/2,0);
        \filldraw[thick, gray!20] (1,0) ellipse (0.8cm and 0.25cm);
        \draw[thick, black] (1,0) ellipse (0.8cm and 0.25cm);
        \filldraw[thick, white] (0.75,0) ellipse (0.45cm and 0.15cm);
        \draw[thick, black] (0.75,0) ellipse (0.45cm and 0.15cm);
         \draw[thick] (A) -- (B) -- (C) -- (D);
        \fill[black] (A) circle (2pt);
        \fill[black] (B) circle (2pt);
        \fill[black] (C) circle (2pt);
        \fill[black] (D) circle (2pt);
    \end{tikzpicture}};.
    \end{align}
Finally, we define the notion of {\it regional tubing} $\rho=\{r_1,\ldots,r_{|\rho|}\}$ of the graph $G$ as a subset of mutually compatible regions for which: 
\begin{itemize}
\item for every region $r \in \rho$ there either exists no sub-regions of $r$ in $\rho$,
\item or $r$ is covered by a collection of sub-regions in $\rho$.
\end{itemize}
We denote the set of tubes appearing as borders between the regions of $\rho$ as $B(\rho)$. For example, the following collection of regions
\begin{equation}
\rho_v =\left\{ \tikz[baseline,yshift=0.5ex] \node[anchor=base] at (0,0) {\begin{tikzpicture}[scale=0.8]
        \coordinate (A) at (0,0);
        \coordinate (B) at (1/2,0);
        \coordinate (C) at (1,0);
        \coordinate (D) at (3/2,0);
        \fillroundrectangle{0}{0.8}{0}{2.2}{0.62}{0};
         \roundrectangle{0}{0.8}{0}{2.2}{0.62}{0};
         \draw[thick] (A) -- (B) -- (C) -- (D);
        \fill[black] (A) circle (2pt);
        \fill[black] (B) circle (2pt);
        \fill[black] (C) circle (2pt);
        \fill[black] (D) circle (2pt);
    \end{tikzpicture}};,\tikz[baseline,yshift=0.5ex] \node[anchor=base] at (0,0) {\begin{tikzpicture}[scale=0.8]
        \coordinate (A) at (0,0);
        \coordinate (B) at (1/2,0);
        \coordinate (C) at (1,0);
        \coordinate (D) at (3/2,0);
        \fillroundrectangle{0}{0.8}{0}{2.2}{0.62}{0};
         \roundrectangle{0}{0.8}{0}{2.2}{0.62}{0};
        \filldraw[thick, white] (0.25,0) ellipse (0.4cm and 0.17cm);
        \draw[thick, black] (0.25,0) ellipse (0.4cm and 0.17cm);
         \draw[thick] (A) -- (B) -- (C) -- (D);
        \fill[black] (A) circle (2pt);
        \fill[black] (B) circle (2pt);
        \fill[black] (C) circle (2pt);
        \fill[black] (D) circle (2pt);
    \end{tikzpicture}};, 
    \tikz[baseline,yshift=0.5ex] \node[anchor=base] at (0,0) {\begin{tikzpicture}[scale=0.8]
        \coordinate (A) at (0,0);
        \coordinate (B) at (1/2,0);
        \coordinate (C) at (1,0);
        \coordinate (D) at (3/2,0);
        \filldraw[thick, gray!20] (0.25,0) ellipse (0.4cm and 0.17cm);
        \draw[thick, black] (0.25,0) ellipse (0.4cm and 0.17cm);
         \draw[thick] (A) -- (B) -- (C) -- (D);
        \fill[black] (A) circle (2pt);
        \fill[black] (B) circle (2pt);
        \fill[black] (C) circle (2pt);
        \fill[black] (D) circle (2pt);
    \end{tikzpicture}};  \tikz[baseline,yshift=0.5ex] \node[anchor=base] at (0,0) {\begin{tikzpicture}[scale=0.8]
        \coordinate (A) at (0,0);
        \coordinate (B) at (1/2,0);
        \coordinate (C) at (1,0);
        \coordinate (D) at (3/2,0);
        \filldraw[thick, gray!20] (0.25,0) ellipse (0.5cm and 0.22cm);
        \draw[thick, black] (0.25,0) ellipse (0.5cm and 0.22cm);
        \filldraw[white] (A) circle (4pt);
        \draw[black,thick] (A) circle (4pt);
        \draw[thick] (A) -- (B) -- (C) -- (D);
        \fill[black] (A) circle (2pt);
        \fill[black] (B) circle (2pt);
        \fill[black] (C) circle (2pt);
        \fill[black] (D) circle (2pt);
    \end{tikzpicture}};,
    \tikz[baseline,yshift=0.5ex] \node[anchor=base] at (0,0) {\begin{tikzpicture}[scale=0.8]
        \coordinate (A) at (0,0);
        \coordinate (B) at (1/2,0);
        \coordinate (C) at (1,0);
        \coordinate (D) at (3/2,0);
        \filldraw[gray!20] (A) circle (4pt);
        \draw[black,thick] (A) circle (4pt);
        \draw[thick] (A) -- (B) -- (C) -- (D);
        \fill[black] (A) circle (2pt);
        \fill[black] (B) circle (2pt);
        \fill[black] (C) circle (2pt);
        \fill[black] (D) circle (2pt);
    \end{tikzpicture}};  \tikz[baseline,yshift=0.5ex] \node[anchor=base] at (0,0) {\begin{tikzpicture}[scale=0.8]
        \coordinate (A) at (0,0);
        \coordinate (B) at (1/2,0);
        \coordinate (C) at (1,0);
        \coordinate (D) at (3/2,0);
        \fillroundrectangle{0}{0.8}{0}{2.2}{0.62}{0};
         \roundrectangle{0}{0.8}{0}{2.2}{0.62}{0};
        \filldraw[thick, white] (.25,0) ellipse (0.45cm and 0.2cm);
        \draw[thick, black] (.25,0) ellipse (0.45cm and 0.2cm);
        \filldraw[white] (D) circle (4pt);
        \draw[black,thick] (D) circle (4pt);
        \draw[thick] (A) -- (B) -- (C) -- (D);
        \fill[black] (A) circle (2pt);
        \fill[black] (B) circle (2pt);
        \fill[black] (C) circle (2pt);
        \fill[black] (D) circle (2pt);
    \end{tikzpicture}};,
    \tikz[baseline,yshift=0.5ex] \node[anchor=base] at (0,0) {\begin{tikzpicture}[scale=0.8]
        \coordinate (A) at (0,0);
        \coordinate (B) at (1/2,0);
        \coordinate (C) at (1,0);
        \coordinate (D) at (3/2,0);
        \filldraw[gray!20] (D) circle (4pt);
        \draw[black,thick] (D) circle (4pt);
        \draw[thick] (A) -- (B) -- (C) -- (D);
        \fill[black] (A) circle (2pt);
        \fill[black] (B) circle (2pt);
        \fill[black] (C) circle (2pt);
        \fill[black] (D) circle (2pt);
    \end{tikzpicture}};     \right\} \,
    \label{eq:reg_tubing_ex},
\end{equation}
is a regional tubing of the path graph $P_4$ whose set of borders is given by
\begin{align}
B\left( \rho_v \right)=\left\{ \tikz[baseline,yshift=0.5ex] \node[anchor=base] at (0,0) {\begin{tikzpicture}[scale=0.8]
        \coordinate (A) at (0,0);
        \coordinate (B) at (1/2,0);
        \coordinate (C) at (1,0);
        \coordinate (D) at (3/2,0);
        \filldraw[thick, white] (0.25,0) ellipse (0.4cm and 0.17cm);
        \draw[thick, black] (0.25,0) ellipse (0.4cm and 0.17cm);
         \draw[thick] (A) -- (B) -- (C) -- (D);
        \fill[black] (A) circle (2pt);
        \fill[black] (B) circle (2pt);
        \fill[black] (C) circle (2pt);
        \fill[black] (D) circle (2pt);
    \end{tikzpicture}};, 
     \tikz[baseline,yshift=0.5ex] \node[anchor=base] at (0,0) {\begin{tikzpicture}[scale=0.8]
        \coordinate (A) at (0,0);
        \coordinate (B) at (1/2,0);
        \coordinate (C) at (1,0);
        \coordinate (D) at (3/2,0);
        \draw[black,thick] (D) circle (4pt);
        \draw[thick] (A) -- (B) -- (C) -- (D);
        \fill[black] (A) circle (2pt);
        \fill[black] (B) circle (2pt);
        \fill[black] (C) circle (2pt);
        \fill[black] (D) circle (2pt);
    \end{tikzpicture}};,\tikz[baseline,yshift=0.5ex] \node[anchor=base] at (0,0) {\begin{tikzpicture}[scale=0.8]
        \coordinate (A) at (0,0);
        \coordinate (B) at (1/2,0);
        \coordinate (C) at (1,0);
        \coordinate (D) at (3/2,0);
        \draw[black,thick] (A) circle (4pt);
        \draw[thick] (A) -- (B) -- (C) -- (D);
        \fill[black] (A) circle (2pt);
        \fill[black] (B) circle (2pt);
        \fill[black] (C) circle (2pt);
        \fill[black] (D) circle (2pt);
    \end{tikzpicture}};     \right\}.
\end{align}
The sub-region relation on the set of regions contained in $\rho_v$ is displayed in Fig.~\ref{fig:reg_poset_exam}. 

A regional tubing $\rho$ is said to refine another regional tubing $\rho'$ if $\rho' \subset \rho$. The set of regional tubings together with inclusion defines a partially ordered set which, as we will argue in the following section, gives the boundary stratification of the {\it graph cosmohedron} $\mathcal{C}_G$. All boundaries of the graph cosmohedron $\mathcal{C}_G$ are then labelled by regional tubings, with the vertices labelled by maximal regional tubings, i.e.~the regional tubings that are maximal elements with respect to inclusion. We will denote the set of all regional tubings by $\Phi_G$, and the set of maximal regional tubings by $\Phi_G^{\text{max}}$.

There exists an alternative diagrammatic representation of regional tubings which arises by endowing tubings with an integer label. Let $\rho$ be a regional tubing and consider the poset on the elements of $\rho$ defined by the partial order relation: $r_1\prec_{\text{reg}} r_2$ if $r_1$ is a sub-region of $r_2$. We define the depth $w$ of a region $r$ in this poset as the maximal length of a chain $r=r_{w} \prec_{\text{reg}} r_{w-1} \prec_{\text{reg}} \ldots \prec_{\text{reg}} r_{0}=r_{G}$\footnote{Here $r_G= \{ V_G \}$ refers to the region corresponding to the root which is the maximal element of the poset defined by the sub-region relation.}. For all regions in $\rho$ of a fixed depth $w$ we take the union of their pairwise borders and assign each element the label $w$. This results in a tubing where each tube receives an integer label. Note, the region $r_G$ associated to the root trivially appears by itself at depth $\textcolor{gray}{0}$ for all regional tubings. To encode this information on a diagram we assign a colour to each label $\{\textcolor{gray}{0}, \textcolor{red}{1}, \textcolor{blue}{2}, \textcolor{green}{3},\ldots \}$ and fill in each tube with the corresponding colour. To illustrate this alternative notation, we consider the following regional tubing that corresponds to a facet of the cosmohedron on $P_4$
\begin{align}
\rho_f=\left\{ \tikz[baseline,yshift=0.5ex] \node[anchor=base] at (0,0) {\begin{tikzpicture}
        \coordinate (A) at (0,0);
        \coordinate (B) at (1/2,0);
        \coordinate (C) at (1,0);
        \coordinate (D) at (3/2,0);
        \fillroundrectangle{0}{0.8}{0}{2.2}{0.62}{0};
         \roundrectangle{0}{0.8}{0}{2.2}{0.62}{0};
         \draw[thick] (A) -- (B) -- (C) -- (D);
        \fill[black] (A) circle (2pt);
        \fill[black] (B) circle (2pt);
        \fill[black] (C) circle (2pt);
        \fill[black] (D) circle (2pt);
    \end{tikzpicture}};,\tikz[baseline,yshift=0.5ex] \node[anchor=base] at (0,0) {\begin{tikzpicture}
        \coordinate (A) at (0,0);
        \coordinate (B) at (1/2,0);
        \coordinate (C) at (1,0);
        \coordinate (D) at (3/2,0);
        \fillroundrectangle{0}{0.8}{0}{2.2}{0.62}{0};
         \roundrectangle{0}{0.8}{0}{2.2}{0.62}{0};
        \filldraw[thick, white] (0.25,0) ellipse (0.4cm and 0.17cm);
        \draw[thick, black] (0.25,0) ellipse (0.4cm and 0.17cm);
         \draw[thick] (A) -- (B) -- (C) -- (D);
        \fill[black] (A) circle (2pt);
        \fill[black] (B) circle (2pt);
        \fill[black] (C) circle (2pt);
        \fill[black] (D) circle (2pt);
    \end{tikzpicture}};, 
    \tikz[baseline,yshift=0.5ex] \node[anchor=base] at (0,0) {\begin{tikzpicture}
        \coordinate (A) at (0,0);
        \coordinate (B) at (1/2,0);
        \coordinate (C) at (1,0);
        \coordinate (D) at (3/2,0);
        \filldraw[thick, gray!20] (0.25,0) ellipse (0.4cm and 0.17cm);
        \draw[thick, black] (0.25,0) ellipse (0.4cm and 0.17cm);
         \draw[thick] (A) -- (B) -- (C) -- (D);
        \fill[black] (A) circle (2pt);
        \fill[black] (B) circle (2pt);
        \fill[black] (C) circle (2pt);
        \fill[black] (D) circle (2pt);
    \end{tikzpicture}};   \right\} = \tikz[baseline,yshift=0.5ex] \node[anchor=base] at (0,0) {\begin{tikzpicture}
        \coordinate (A) at (0,0);
        \coordinate (B) at (1/2,0);
        \coordinate (C) at (1,0);
        \coordinate (D) at (3/2,0);
        \fillroundrectangle{0}{0.8}{0}{2.2}{0.62}{0};
         \roundrectangle{0}{0.8}{0}{2.2}{0.62}{0};
        \filldraw[thick, red!20] (0.25,0) ellipse (0.4cm and 0.17cm);
        \draw[thick, black] (0.25,0) ellipse (0.4cm and 0.17cm);
        \draw[thick] (A) -- (B) -- (C) -- (D);
        \fill[black] (A) circle (2pt);
        \fill[black] (B) circle (2pt);
        \fill[black] (C) circle (2pt);
        \fill[black] (D) circle (2pt);
    \end{tikzpicture}};.
\end{align}
A co-dimension two boundary i.e.~an edge of the same cosmohedron is 
\begin{align}
\rho_e= \rho_f \cup \left\{ 
    \tikz[baseline,yshift=0.5ex] \node[anchor=base] at (0,0) {\begin{tikzpicture}
        \coordinate (A) at (0,0);
        \coordinate (B) at (1/2,0);
        \coordinate (C) at (1,0);
        \coordinate (D) at (3/2,0);
        \filldraw[thick, gray!20] (0.25,0) ellipse (0.5cm and 0.22cm);
        \draw[thick, black] (0.25,0) ellipse (0.5cm and 0.22cm);
         \draw[thick] (A) -- (B) -- (C) -- (D);
        \filldraw[white] (A) circle (4pt);
        \draw[black,thick] (A) circle (4pt);
        \fill[black] (A) circle (2pt);
        \fill[black] (B) circle (2pt);
        \fill[black] (C) circle (2pt);
        \fill[black] (D) circle (2pt);
    \end{tikzpicture}};,
    \tikz[baseline,yshift=0.5ex] \node[anchor=base] at (0,0) {\begin{tikzpicture}
        \coordinate (A) at (0,0);
        \coordinate (B) at (1/2,0);
        \coordinate (C) at (1,0);
        \coordinate (D) at (3/2,0);
         \draw[thick] (A) -- (B) -- (C) -- (D);
        \filldraw[gray!20] (A) circle (4pt);
        \draw[black,thick] (A) circle (4pt);
        \fill[black] (A) circle (2pt);
        \fill[black] (B) circle (2pt);
        \fill[black] (C) circle (2pt);
        \fill[black] (D) circle (2pt);
    \end{tikzpicture}};    \right\} = \tikz[baseline,yshift=0.5ex] \node[anchor=base] at (0,0) {\begin{tikzpicture}
        \coordinate (A) at (0,0);
        \coordinate (B) at (1/2,0);
        \coordinate (C) at (1,0);
        \coordinate (D) at (3/2,0);
        \fillroundrectangle{0}{0.8}{0}{2.2}{0.62}{0};
         \roundrectangle{0}{0.8}{0}{2.2}{0.62}{0};
        \filldraw[thick, red!20] (0.25,0) ellipse (0.5cm and 0.22cm);
        \draw[thick, black] (0.25,0) ellipse (0.5cm and 0.22cm);
        \filldraw[blue!25] (A) circle (4pt);
        \draw[black,thick] (A) circle (4pt);
        \draw[thick] (A) -- (B) -- (C) -- (D);
        \fill[black] (A) circle (2pt);
        \fill[black] (B) circle (2pt);
        \fill[black] (C) circle (2pt);
        \fill[black] (D) circle (2pt);
    \end{tikzpicture}};,
\end{align}
while a vertex is given by
\begin{align}
\rho_{v}= \rho_e \cup \left\{ 
    \tikz[baseline,yshift=0.5ex] \node[anchor=base] at (0,0) {\begin{tikzpicture}
        \coordinate (A) at (0,0);
        \coordinate (B) at (1/2,0);
        \coordinate (C) at (1,0);
        \coordinate (D) at (3/2,0);
        \fillroundrectangle{0}{0.8}{0}{2.2}{0.62}{0};
         \roundrectangle{0}{0.8}{0}{2.2}{0.62}{0};
        \filldraw[thick, white] (.25,0) ellipse (0.45cm and 0.2cm);
        \draw[thick, black] (.25,0) ellipse (0.45cm and 0.2cm);
         \draw[thick] (A) -- (B) -- (C) -- (D);
        \filldraw[white] (D) circle (4pt);
        \draw[black,thick] (D) circle (4pt);
        \fill[black] (A) circle (2pt);
        \fill[black] (B) circle (2pt);
        \fill[black] (C) circle (2pt);
        \fill[black] (D) circle (2pt);
    \end{tikzpicture}};,
    \tikz[baseline,yshift=0.5ex] \node[anchor=base] at (0,0) {\begin{tikzpicture}
        \coordinate (A) at (0,0);
        \coordinate (B) at (1/2,0);
        \coordinate (C) at (1,0);
        \coordinate (D) at (3/2,0);
         \draw[thick] (A) -- (B) -- (C) -- (D);
        \filldraw[gray!20] (D) circle (4pt);
        \draw[black,thick] (D) circle (4pt);
        \fill[black] (A) circle (2pt);
        \fill[black] (B) circle (2pt);
        \fill[black] (C) circle (2pt);
        \fill[black] (D) circle (2pt);
    \end{tikzpicture}};    \right\} = \tikz[baseline,yshift=0.5ex] \node[anchor=base] at (0,0) {\begin{tikzpicture}
        \coordinate (A) at (0,0);
        \coordinate (B) at (1/2,0);
        \coordinate (C) at (1,0);
        \coordinate (D) at (3/2,0);
        \fillroundrectangle{0}{0.8}{0}{2.2}{0.62}{0};
         \roundrectangle{0}{0.8}{0}{2.2}{0.62}{0};
        \filldraw[thick, red!20] (0.25,0) ellipse (0.5cm and 0.22cm);
        \draw[thick, black] (0.25,0) ellipse (0.5cm and 0.22cm);
        \filldraw[blue!25] (A) circle (4pt);
        \draw[black,thick] (A) circle (4pt);
        \filldraw[blue!25] (D) circle (4pt);
        \draw[black,thick] (D) circle (4pt);
        \draw[thick] (A) -- (B) -- (C) -- (D);
        \fill[black] (A) circle (2pt);
        \fill[black] (B) circle (2pt);
        \fill[black] (C) circle (2pt);
        \fill[black] (D) circle (2pt);
    \end{tikzpicture}};.
    \label{eq:reg_tube_exam}
\end{align} 
To better understand this alternative notation, one can compare this representation of $\rho_v$ with the one in Fig.~\ref{fig:reg_poset_exam}. In this case the only tube appearing as a border at level $\textcolor{red}{1}$ is the tube $\{ 1,2\}$. At level two, there are two tubes which appear as borders, namely the tubes $\{1\}$ and $\{4\}$. For more complicated graphs these diagrams no longer serve the purpose of simplifying notation, in which case we revert back to the definition of regional tubings in terms of either sets of regions or tubings endowed with the additional integer labels, as previously described.

It is worth emphasising that generally the partial ordering of regional tubings described above will lead to boundary posets for non-simple polytopes, as already noted in \cite{Arkani-Hamed:2024jbp}. For example, the regional tubing $\rho_v$ from \eqref{eq:reg_tube_exam} corresponds to a non-simple vertex of the cosmohedron on $P_4$ which sits in the boundary of four facets. We will return to this point in detail in section \ref{sec:can_form}.

\begin{figure}
\center
\tikz[baseline,yshift=0.5ex] \node[anchor=base] at (0,0) {\begin{tikzpicture}[scale=1.1]
        \coordinate (A) at (0,0);
        \coordinate (B) at (1/2,0);
        \coordinate (C) at (1,0);
        \coordinate (D) at (3/2,0);
        \fillroundrectangle{0}{0.8}{0}{2.2}{0.62}{0};
         \roundrectangle{0}{0.8}{0}{2.2}{0.62}{0};
         \draw[thick] (A) -- (B) -- (C) -- (D);
        \fill[black] (A) circle (2pt);
        \fill[black] (B) circle (2pt);
        \fill[black] (C) circle (2pt);
        \fill[black] (D) circle (2pt);
        \coordinate (A) at (0-3.5,0-1.5);
        \coordinate (B) at (1/2-3.5,0-1.5);
        \coordinate (C) at (1-3.5,0-1.5);
        \coordinate (D) at (3/2-3.5,0-1.5);
        \fillroundrectangle{0}{0.8-3.5}{0-1.5}{2.2}{0.62}{0};
         \roundrectangle{0}{0.8-3.5}{0-1.5}{2.2}{0.62}{0};
        \filldraw[thick, white] (0.25-3.5,0-1.5) ellipse (0.5cm and 0.15cm);
        \draw[thick, black] (0.25-3.5,0-1.5) ellipse (0.5cm and 0.15cm);
         \draw[thick] (A) -- (B) -- (C) -- (D);
        \fill[black] (A) circle (2pt);
        \fill[black] (B) circle (2pt);
        \fill[black] (C) circle (2pt);
        \fill[black] (D) circle (2pt);
        \coordinate (A) at (0-3.5-1.5,0-3);
        \coordinate (B) at (1/2-3.5-1.5,0-3);
        \coordinate (C) at (1-3.5-1.5,0-3);
        \coordinate (D) at (3/2-3.5-1.5,0-3);
        \fillroundrectangle{0}{0.8-3.5-1.5}{0-3}{2.2}{0.62}{0};
         \roundrectangle{0}{0.8-3.5-1.5}{0-3}{2.2}{0.62}{0};
        \filldraw[thick, white] (0.25-3.5-1.5,0-3) ellipse (0.5cm and 0.22cm);
        \draw[thick, black] (0.25-3.5-1.5,0-3) ellipse (0.5cm and 0.22cm);
        \fill[white] (D) circle (4pt);
        \draw[thick] (D) circle (4pt);
         \draw[thick] (A) -- (B) -- (C) -- (D);
        \fill[black] (A) circle (2pt);
        \fill[black] (B) circle (2pt);
        \fill[black] (C) circle (2pt);
        \fill[black] (D) circle (2pt);
        \coordinate (A) at (0-3.5+1.5,0-3);
        \coordinate (B) at (1/2-3.5+1.5,0-3);
        \coordinate (C) at (1-3.5+1.5,0-3);
        \coordinate (D) at (3/2-3.5+1.5,0-3);
        \fill[gray!20] (D) circle (4pt);
        \draw[thick] (D) circle (4pt);
         \draw[thick] (A) -- (B) -- (C) -- (D);
        \fill[black] (A) circle (2pt);
        \fill[black] (B) circle (2pt);
        \fill[black] (C) circle (2pt);
        \fill[black] (D) circle (2pt);
        \coordinate (A) at (0+3.5,0-1.5);
        \coordinate (B) at (1/2+3.5,0-1.5);
        \coordinate (C) at (1+3.5,0-1.5);
        \coordinate (D) at (3/2+3.5,0-1.5);
        \filldraw[thick, gray!20] (0.25+3.5,0-1.5) ellipse (0.5cm and 0.15cm);
        \draw[thick, black] (0.25+3.5,0-1.5) ellipse (0.5cm and 0.15cm);
         \draw[thick] (A) -- (B) -- (C) -- (D);
        \fill[black] (A) circle (2pt);
        \fill[black] (B) circle (2pt);
        \fill[black] (C) circle (2pt);
        \fill[black] (D) circle (2pt);
        \coordinate (A) at (0+3.5-1.5,0-3);
        \coordinate (B) at (1/2+3.5-1.5,0-3);
        \coordinate (C) at (1+3.5-1.5,0-3);
        \coordinate (D) at (3/2+3.5-1.5,0-3);
        \filldraw[thick, gray!20] (0.25+3.5-1.5,0-3) ellipse (0.5cm and 0.22cm);
        \draw[thick, black] (0.25+3.5-1.5,0-3) ellipse (0.5cm and 0.22cm);
        \fill[white] (A) circle (4pt);
        \draw[thick] (A) circle (4pt);
         \draw[thick] (A) -- (B) -- (C) -- (D);
        \fill[black] (A) circle (2pt);
        \fill[black] (B) circle (2pt);
        \fill[black] (C) circle (2pt);
        \fill[black] (D) circle (2pt);
        \coordinate (A) at (0+3.5+1.5,0-3);
        \coordinate (B) at (1/2+3.5+1.5,0-3);
        \coordinate (C) at (1+3.5+1.5,0-3);
        \coordinate (D) at (3/2+3.5+1.5,0-3);
        \fill[gray!20] (A) circle (4pt);
        \draw[thick] (A) circle (4pt);
         \draw[thick] (A) -- (B) -- (C) -- (D);
        \fill[black] (A) circle (2pt);
        \fill[black] (B) circle (2pt);
        \fill[black] (C) circle (2pt);
        \fill[black] (D) circle (2pt);
        \draw[thick] (0.5,-0.5)--(-3+1-0.75,-1.);
        \draw[thick] (1,-0.5)--(+3+0.5+0.75,-1.);
        \draw[thick] (-3+1-0.5,-2.5+0.5) -- (-3+1+0.75,-2.5);
        \draw[thick] (-3+1-1,-2.5+0.5) -- (-3+1+0.75-3,-2.5);
        \draw[thick] (3+1,-2.5+0.5) -- (3-0.25,-2.5);
         \draw[thick] (3+1.5,-2.5+0.5) -- (3+1.5+1.25,-2.5);
    \end{tikzpicture}};
    \caption{The sub-region relation on the set of regions contained in the regional tubing $\rho_v$ defined in \eqref{eq:reg_tubing_ex}. This corresponds to a non-simple vertex of the corresponding cosmohedron for the graph $P_4$.}
    \label{fig:reg_poset_exam}
\end{figure}

\subsubsection{Boundaries and factorisation}
\begin{figure}
\center
\begin{tikzpicture}[scale=0.7]
	\coordinate (A) at (0,-1.5);
        \coordinate (B) at (1/2,-1.5);
        \coordinate (C) at (1,-1.5);
        \coordinate (D) at (3/2,-1.5);
        \fillroundrectangle{0}{0.8}{-1.5}{2.2}{0.62}{0};
         \roundrectangle{0}{0.8}{-1.5}{2.2}{0.62}{0};
        \filldraw[red!25] (A) circle (4pt);
        \draw[black,thick] (A) circle (4pt);
        \filldraw[red!25] (C) circle (4pt);
        \draw[black,thick] (C) circle (4pt);
        \draw[thick] (A) -- (B) -- (C) -- (D);
        \fill[black] (A) circle (2pt);
        \fill[black] (B) circle (2pt);
        \fill[black] (C) circle (2pt);
        \fill[black] (D) circle (2pt);
	\coordinate (A) at (0-4.5,-1.5);
        \coordinate (B) at (1/2-4.5,-1.5);
        \coordinate (C) at (1-4.5,-1.5);
        \coordinate (D) at (3/2-4.5,-1.5);
        \fillroundrectangle{0}{0.8-4.5}{-1.5}{2.2}{0.62}{0};
         \roundrectangle{0}{0.8-4.5}{-1.5}{2.2}{0.62}{0};
        \filldraw[red!25] (A) circle (4pt);
        \draw[black,thick] (A) circle (4pt);
        \filldraw[thick, blue!25] (1.25-4.5,-1.5) ellipse (0.5cm and 0.25cm);
        \draw[thick, black] (1.25-4.5,-1.5) ellipse (0.5cm and 0.25cm);
        \filldraw[red!25] (C) circle (4pt);
        \draw[black,thick] (C) circle (4pt);
        \draw[thick] (A) -- (B) -- (C) -- (D);
        \fill[black] (A) circle (2pt);
        \fill[black] (B) circle (2pt);
        \fill[black] (C) circle (2pt);
        \fill[black] (D) circle (2pt);
	\coordinate (A) at (0+4.5,-1.5);
        \coordinate (B) at (1/2+4.5,-1.5);
        \coordinate (C) at (1+4.5,-1.5);
        \coordinate (D) at (3/2+4.5,-1.5);
        \fillroundrectangle{0}{0.8+4.5}{-1.5}{2.2}{0.62}{0};
         \roundrectangle{0}{0.8+4.5}{-1.5}{2.2}{0.62}{0};
        \filldraw[thick, blue!20] (0.5+4.5,-1.5) ellipse (0.75cm and 0.25cm);
        \draw[thick, black] (0.5+4.5,-1.5) ellipse (0.75cm and 0.25cm);
         \filldraw[red!25] (A) circle (4pt);
        \draw[black,thick] (A) circle (4pt);
        \filldraw[red!25] (C) circle (4pt);
        \draw[black,thick] (C) circle (4pt);
        \draw[thick] (A) -- (B) -- (C) -- (D);
        \fill[black] (A) circle (2pt);
        \fill[black] (B) circle (2pt);
        \fill[black] (C) circle (2pt);
        \fill[black] (D) circle (2pt);
	\coordinate (A) at (0+3+1.5,-1.5+3.5);
        \coordinate (B) at (1/2+3+1.5,-1.5+3.5);
        \coordinate (C) at (1+3+1.5,-1.5+3.5);
        \coordinate (D) at (3/2+3+1.5,-1.5+3.5);
        \fillroundrectangle{0}{0.8+3+1.5}{-1.5+3.5}{2.2}{0.62}{0};
         \roundrectangle{0}{0.8+3+1.5}{-1.5+3.5}{2.2}{0.62}{0};
         \filldraw[thick, green!30] (0.5+3+1.5,-1.5+3.5) ellipse (0.75cm and 0.25cm);
        \draw[thick, black] (0.5+3+1.5,-1.5+3.5) ellipse (0.75cm and 0.25cm);
        \filldraw[blue!25] (A) circle (4pt);
        \draw[black,thick] (A) circle (4pt);
        \filldraw[red!25] (C) circle (4pt);
        \draw[black,thick] (C) circle (4pt);
        \draw[thick] (A) -- (B) -- (C) -- (D);
        \fill[black] (A) circle (2pt);
        \fill[black] (B) circle (2pt);
        \fill[black] (C) circle (2pt);
        \fill[black] (D) circle (2pt);
	\coordinate (A) at (0-4.5,-1.5+3.5);
        \coordinate (B) at (1/2-4.5,-1.5+3.5);
        \coordinate (C) at (1-4.5,-1.5+3.5);
        \coordinate (D) at (3/2-4.5,-1.5+3.5);
        \fillroundrectangle{0}{0.8-4.5}{-1.5+3.5}{2.2}{0.62}{0};
         \roundrectangle{0}{0.8-4.5}{-1.5+3.5}{2.2}{0.62}{0};
         \filldraw[thick, green!30] (1.25-4.5,-1.5+3.5) ellipse (0.5cm and 0.25cm);
        \draw[thick, black] (1.25-4.5,-1.5+3.5) ellipse (0.5cm and 0.25cm);
        \filldraw[blue!25] (A) circle (4pt);
        \draw[black,thick] (A) circle (4pt);
        \filldraw[red!25] (C) circle (4pt);
        \draw[black,thick] (C) circle (4pt);
        \draw[thick] (A) -- (B) -- (C) -- (D);
        \fill[black] (A) circle (2pt);
        \fill[black] (B) circle (2pt);
        \fill[black] (C) circle (2pt);
        \fill[black] (D) circle (2pt);
	\coordinate (A) at (0,-1.5+3.5);
        \coordinate (B) at (1/2,-1.5+3.5);
        \coordinate (C) at (1,-1.5+3.5);
        \coordinate (D) at (3/2,-1.5+3.5);
        \fillroundrectangle{0}{0.8}{-1.5+3.5}{2.2}{0.62}{0};
         \roundrectangle{0}{0.8}{-1.5+3.5}{2.2}{0.62}{0};
        \filldraw[blue!25] (A) circle (4pt);
        \draw[black,thick] (A) circle (4pt);
        \filldraw[red!25] (C) circle (4pt);
        \draw[black,thick] (C) circle (4pt);
        \draw[thick] (A) -- (B) -- (C) -- (D);
        \fill[black] (A) circle (2pt);
        \fill[black] (B) circle (2pt);
        \fill[black] (C) circle (2pt);
        \fill[black] (D) circle (2pt);
	\coordinate (A) at (0-4.5,-1.5-3.5);
        \coordinate (B) at (1/2-4.5,-1.5-3.5);
        \coordinate (C) at (1-4.5,-1.5-3.5);
        \coordinate (D) at (3/2-4.5,-1.5-3.5);
        \fillroundrectangle{0}{0.8-4.5}{-1.5-3.5}{2.2}{0.62}{0};
         \roundrectangle{0}{0.8-4.5}{-1.5-3.5}{2.2}{0.62}{0};
          \filldraw[thick, green!30] (1.25-4.5,-1.5-3.5) ellipse (0.5cm and 0.25cm);
        \draw[thick, black] (1.25-4.5,-1.5-3.5) ellipse (0.5cm and 0.25cm);
        \filldraw[red!25] (A) circle (4pt);
        \draw[black,thick] (A) circle (4pt);
        \filldraw[blue!25] (C) circle (4pt);
        \draw[black,thick] (C) circle (4pt);
        \draw[thick] (A) -- (B) -- (C) -- (D);
        \fill[black] (A) circle (2pt);
        \fill[black] (B) circle (2pt);
        \fill[black] (C) circle (2pt);
        \fill[black] (D) circle (2pt);
	\coordinate (A) at (0+4.5,-1.5-3.5);
        \coordinate (B) at (1/2+4.5,-1.5-3.5);
        \coordinate (C) at (1+4.5,-1.5-3.5);
        \coordinate (D) at (3/2+4.5,-1.5-3.5);
        \fillroundrectangle{0}{0.8+4.5}{-1.5-3.5}{2.2}{0.62}{0};
         \roundrectangle{0}{0.8+4.5}{-1.5-3.5}{2.2}{0.62}{0};
         \filldraw[thick, green!30] (0.5+4.5,-1.5-3.5) ellipse (0.75cm and 0.25cm);
        \draw[thick, black] (0.5+4.5,-1.5-3.5) ellipse (0.75cm and 0.25cm);
        \filldraw[red!25] (A) circle (4pt);
        \draw[black,thick] (A) circle (4pt);
        \filldraw[blue!25] (C) circle (4pt);
        \draw[black,thick] (C) circle (4pt);
        \draw[thick] (A) -- (B) -- (C) -- (D);
        \fill[black] (A) circle (2pt);
        \fill[black] (B) circle (2pt);
        \fill[black] (C) circle (2pt);
        \fill[black] (D) circle (2pt);
	\coordinate (A) at (0,-1.5-3.5);
        \coordinate (B) at (1/2,-1.5-3.5);
        \coordinate (C) at (1,-1.5-3.5);
        \coordinate (D) at (3/2,-1.5-3.5);
        \fillroundrectangle{0}{0.8}{-1.5-3.5}{2.2}{0.62}{0};
         \roundrectangle{0}{0.8}{-1.5-3.5}{2.2}{0.62}{0};
        \filldraw[red!25] (A) circle (4pt);
        \draw[black,thick] (A) circle (4pt);
        \filldraw[blue!25] (C) circle (4pt);
        \draw[black,thick] (C) circle (4pt);
        \draw[thick] (A) -- (B) -- (C) -- (D);
        \fill[black] (A) circle (2pt);
        \fill[black] (B) circle (2pt);
        \fill[black] (C) circle (2pt);
        \fill[black] (D) circle (2pt);
        \draw[thick] (3.75,1.5)--(0.75-3,1.5);
        \draw[thick] (3.75,-4.5)--(0.75-3,-4.5);
        \draw[thick] (3.75,1.5)--(3.75,-4.5);
        \draw[thick] (-2.25,1.5)--(-2.25,-4.5);
        \fill[black] (0.75+3,1.5) circle (2pt);
         \fill[black] (0.75-3,1.5) circle (2pt);
         \fill[black] (0.75+3,-4.5) circle (2pt);
          \fill[black] (0.75-3,-4.5) circle (2pt);
\end{tikzpicture}
\caption{An illustration of the factorisation property of a facet of the cosmohedron for the graph $P_4$ with labels ordered as $\{\textcolor{gray}{0},\textcolor{red}{1},\textcolor{blue}{2},\textcolor{green}{3}\}$. }
\label{fig:fac}
\end{figure}
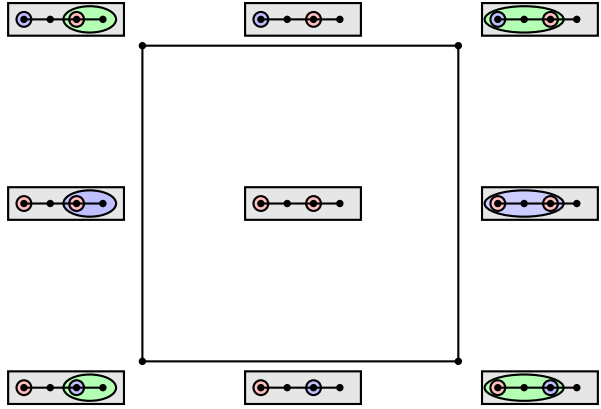

Similar as for graph associahedra, also graph cosmohedra have simple factorisation properties of their facets. To describe them, we introduce the notion of the {\it spine} of a tubing. Given a tubing $\tau$ of a graph $G$ we define a graph $\text{spine}_\tau(G)$ whose vertices are given by the tubes $t \in \tau$ such that two vertices $t_1$ and $t_2$ are connected by an edge if $t_1$ and $t_2$ share the same parent or one is the parent of the other in the graded poset $\mathcal{P}_{\overline{\tau}}(G)$. As an example we have the following spine graphs
\begin{align}
\text{spine} 
\left(
\tikz[baseline,yshift=0.5ex] \node[anchor=base] at (0,0) {\begin{tikzpicture}
        \coordinate (A) at (0,0);
        \coordinate (B) at (1/2,0);
        \coordinate (C) at (1,0);
        \coordinate (D) at (3/2,0);
        \coordinate (E) at (2,0);
        \coordinate (F) at (5/2,0);
         \roundrectangle{0}{1.2}{0}{3.2}{0.62}{0};
        \draw[thick, black] (0.75,0) ellipse (1cm and 0.27cm);
        \draw[thick, black] (0.25,0) ellipse (0.4cm and 0.17cm);
        \draw[thick] (A) -- (B) -- (C) -- (D) -- (E) --(F);
        \fill[black] (A) circle (2pt);
        \fill[black] (B) circle (2pt);
        \fill[black] (C) circle (2pt);
        \fill[black] (D) circle (2pt);
        \fill[black] (E) circle (2pt);
        \fill[black] (F) circle (2pt);
        \draw[thick] (F) circle (4pt);
                \draw[thick] (D) circle (4pt);
    \end{tikzpicture}};
\right) = \tikz[baseline,yshift=4.6ex] \node[anchor=base] at (0,0) {\begin{tikzpicture}
        \coordinate (B) at (-1/4,-1/2);
         \coordinate (C) at (+1/4,-1/2);
         \coordinate (D) at (-1/2,-1);
                  \coordinate (E) at (0,-1);
        \draw[thick] (C) -- (B)--(E);
        \draw[thick] (B) -- (D)--(E);
        \fill[black] (B) circle (2pt);
        \fill[black] (C) circle (2pt);
        \fill[black] (D) circle (2pt);
                \fill[black] (E) circle (2pt);
    \end{tikzpicture}};, \quad 
    \text{spine} 
\left(
\tikz[baseline,yshift=0.5ex] \node[anchor=base] at (0,0) {\begin{tikzpicture}
        \coordinate (A) at (0,0);
        \coordinate (B) at (1/2,0);
        \coordinate (C) at (1,0);
        \coordinate (D) at (3/2,0);
        \coordinate (E) at (2,0);
        \coordinate (F) at (5/2,0);
                 \roundrectangle{0}{1.2}{0}{3.2}{0.62}{0};
        \draw[thick, black] (0.25,0) ellipse (0.5cm and 0.25cm);
        \draw[thick, black] (2.25,0) ellipse (0.5cm and 0.25cm);
        \draw[thick] (A) -- (B) -- (C) -- (D) -- (E) --(F);
        \fill[black] (A) circle (2pt);
        \fill[black] (B) circle (2pt);
        \fill[black] (C) circle (2pt);
        \fill[black] (D) circle (2pt);
        \fill[black] (E) circle (2pt);
        \fill[black] (F) circle (2pt);
        \draw[thick] (F) circle (4pt);
    \end{tikzpicture}};
\right) &= \tikz[baseline,yshift=4.6ex] \node[anchor=base] at (0,0) {\begin{tikzpicture}
        \coordinate (B) at (-1/4,-1/2);
         \coordinate (C) at (+1/4,-1/2);
         \coordinate (D) at (+1/4,-1);
        \draw[thick] (B) -- (C);
        \draw[thick] (C) -- (D);
        \fill[black] (B) circle (2pt);
        \fill[black] (C) circle (2pt);
        \fill[black] (D) circle (2pt);
    \end{tikzpicture}};.
\end{align}
With this definition the factorisation property on the co-dimension one boundaries of the graph cosmohedra, labelled by the tubing $\tau$, are given by 
\begin{align}
\partial_{\tau}(\mathcal{C}_G) = \mathcal{A}_{\text{spine}_\tau(G)} \times \prod_{r \in R(\tau)} \mathcal{C} _{G[t_p^{(r)}]^*_{\{t_{c,i}^{(r)}\}}},
\end{align}
where the latter factor contains the graph cosmohedra for the graphs obtained by restricting $G$ to one of the regions of $\tau$. An example of the factorisation of a co-dimension one boundary of the cosmohedron for the graph $P_4$ is displayed in Fig.~\ref{fig:fac}. In this case we have $\tau=\{\{1\},\{3\}\}$ and $\text{spine}_\tau(P_4)=P_2$. Moreover, there are three regions generated by the tubing $\tau$: two of them result in the trivial reconnected components $P_1$ for which the cosmohedron is a point, and the third one results in the reconnected component $P_2$, for which the cosmohedron is a segment. Therefore, the facet of the cosmohedron $\mathcal{C}_{P_4}$ specified by $\tau$ is the Cartesian product of two segments.
 
\subsection{Cosmological Amplitubes}
Having defined the combinatorics of the graph cosmohedra we can now go ahead and define their corresponding {\it cosmological amplitubes}
as 
\begin{align}
\Psi_{G} = \sum_{\rho\in \Phi_G^{\text{max}}} \prod_{r \in\rho} \frac{1}{\mathcal{R}_r}\,,
\end{align}
where we assigned the variable $\mathcal{R}_r$ to the region $r$. We also set $\mathcal{R}_{V_g}=1$ for the region containing the root as the parent, and no children. This formula correctly reproduces the one found in \cite{Arkani-Hamed:2024jbp} since the set of regional tubing agrees with the set of non-overlapping subpolygons of an $n$-gon for path graphs, and the regions are in one-to-one correspondence with subpolygons.

\begin{figure}[t]
 \centering
 \includegraphics[width=0.7\textwidth]{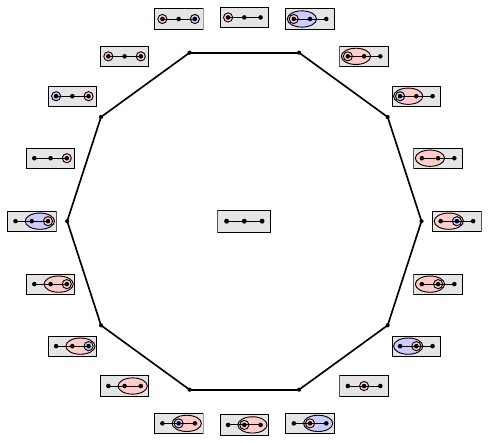}
 \caption{The graph cosmohedron for the path graph on three vertices which reproduces the two dimensional cosmohedron.}
    \label{fig:cosmo_P3_exam}
\end{figure}

As an example we consider again the path graph $P_3$ on three vertices.  In this case the graph cosmohedron $\mathcal{C}_{P_3}$ has $10$ vertices with the corresponding regional tubings
\begin{align}
\rho^{(P_3)}_1 = \raisebox{-0.5ex}{\includegraphics{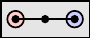}}, \quad \rho^{(P_3)}_2 =\raisebox{-0.5ex}{\includegraphics{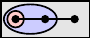}}, \quad \ldots, \quad\rho^{(P_3)}_{10} =\raisebox{-0.5ex}{\includegraphics{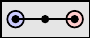}},
\end{align}
where the remaining regional tubings can be read off from Fig.~\ref{fig:cosmo_P3_exam}.
Then, the cosmological amplitube associated to the path graph $P_3$ is\footnote{By convention we have chosen to drop an overall factor coming from the region  $\mathcal{R}_{V_G}$ associated to the entire graph.} 
\begin{align}
\Psi_{P_3} &= \frac{1}{\mathcal{R}_{r_1}\mathcal{R}_{r_5}\mathcal{R}_{r_{7}}\mathcal{R}_{r_{12}}}+\frac{1}{\mathcal{R}_{r_5}\mathcal{R}_{r_7}\mathcal{R}_{r_{10}}\mathcal{R}_{r_{12}}}+\frac{1}{\mathcal{R}_{r_5}\mathcal{R}_{r_8}\mathcal{R}_{r_{10}}\mathcal{R}_{r_{13}}}+\frac{1}{\mathcal{R}_{r_2}\mathcal{R}_{r_5}\mathcal{R}_{r_{8}}\mathcal{R}_{r_{13}}}\nonumber\\&+\frac{1}{\mathcal{R}_{r_2}\mathcal{R}_{r_6}\mathcal{R}_{r_{8}}\mathcal{R}_{r_{14}}}+\frac{1}{\mathcal{R}_{r_6}\mathcal{R}_{r_8}\mathcal{R}_{r_{11}}\mathcal{R}_{r_{14}}}+\frac{1}{\mathcal{R}_{r_6}\mathcal{R}_{r_9}\mathcal{R}_{r_{11}}\mathcal{R}_{r_{15}}}+\frac{1}{\mathcal{R}_{r_3}\mathcal{R}_{r_6}\mathcal{R}_{r_{9}}\mathcal{R}_{r_{15}}}\nonumber\\&+\frac{1}{\mathcal{R}_{r_3}\mathcal{R}_{r_4}\mathcal{R}_{r_{7}}\mathcal{R}_{r_{9}}}+\frac{1}{\mathcal{R}_{r_1}\mathcal{R}_{r_4}\mathcal{R}_{r_{7}}\mathcal{R}_{r_{9}}}\,,
\end{align}
whose terms are in one-to-one correspondence to that of the wavefunction of $\tr(\phi^3)$ theory \cite{Arkani-Hamed:2024jbp}. Note, the expression for the cosmological amplitube can be re-organised into a sum over vertices of the corresponding graph associahedron as
\begin{align}
\Psi_{P_3} &= \frac{1}{\mathcal{R}_{r_5}\mathcal{R}_{r_{7}}\mathcal{R}_{r_{12}}} \left( \frac{1}{\mathcal{R}_{r_1}}+\frac{1}{\mathcal{R}_{r_{10}}}\right)
+\frac{1}{\mathcal{R}_{r_5}\mathcal{R}_{r_{8}}\mathcal{R}_{r_{13}}} \left( \frac{1}{\mathcal{R}_{r_{10}}}+\frac{1}{\mathcal{R}_{r_{2}}}\right)\notag \\
&+\frac{1}{\mathcal{R}_{r_6}\mathcal{R}_{r_{8}}\mathcal{R}_{r_{14}}} \left( \frac{1}{\mathcal{R}_{r_{2}}}+\frac{1}{\mathcal{R}_{r_{11}}}\right)
+\frac{1}{\mathcal{R}_{r_6}\mathcal{R}_{r_{9}}\mathcal{R}_{r_{15}}} \left( \frac{1}{\mathcal{R}_{r_{11}}}+\frac{1}{\mathcal{R}_{r_{3}}}\right) \notag\\
&+\frac{1}{\mathcal{R}_{r_4}\mathcal{R}_{r_{7}}\mathcal{R}_{r_{9}}} \left( \frac{1}{\mathcal{R}_{r_{3}}}+\frac{1}{\mathcal{R}_{r_{1}}}\right).
\end{align}
In the physics literature this decomposition is refered to as a sum over `channels'. Generally, each vertex of the graph associahedron is labelled by a maximal tubing $\tau$ and contributes $|\Gamma_{\text{spine}_{\tau}(G)}|$ many terms to the cosmological amplitube.

\subsection{Cosmohedron Embedding}
We now move on to provide an embedding for the graph cosmohedron for any graph $G$. We will start with the coordinate system that we introduced for the tube embedding of the graph associahedron $\mathcal{A}_G$ described in section \ref{sec:ass_emb}, and provide additional inequalities for each tubing which will carve off facets of this polytope in order to arrive at the corresponding graph cosmohedron $\mathcal{C}_G$. This construction will reproduce the cosmohedra defined in \cite{Arkani-Hamed:2024jbp} for path graphs, but will provide previously unknown polytopes for a general graph. 

\begin{figure}[t]
    \centering
    \includegraphics[width=0.27\textwidth]{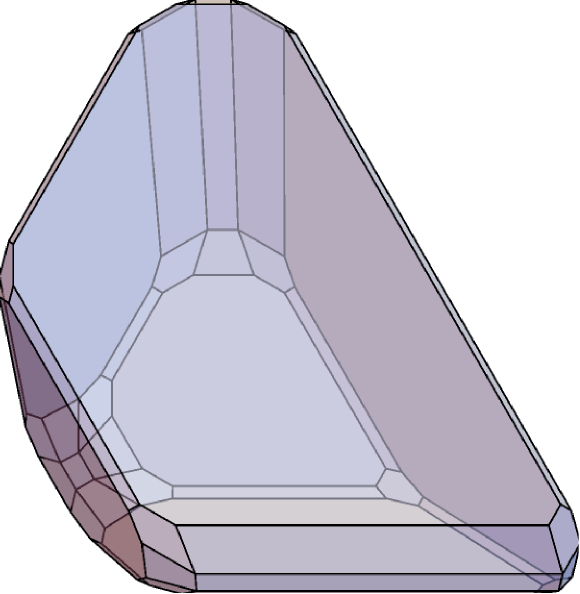} \quad \quad \includegraphics[width=0.31\textwidth]{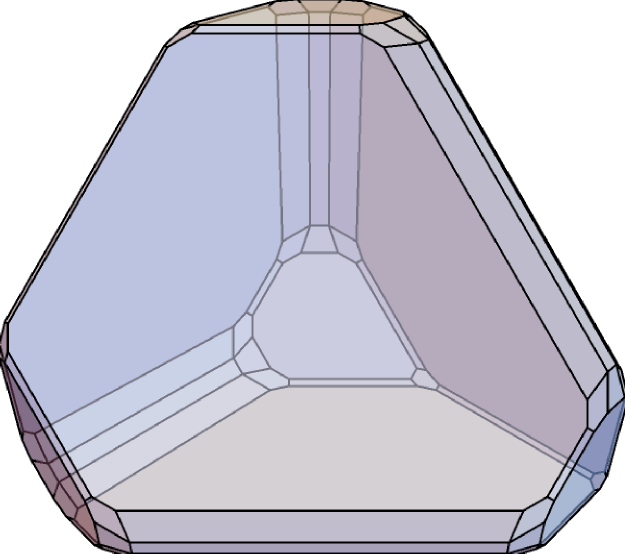} \quad \quad\includegraphics[width=0.3\textwidth]{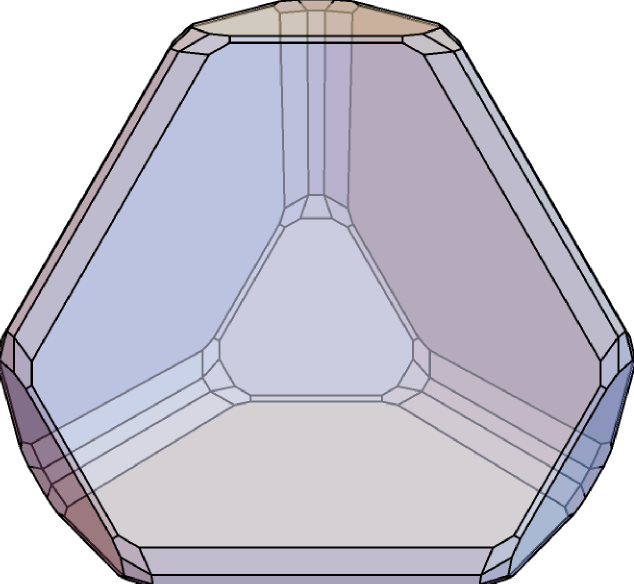}
    \caption{Embeddings of three-dimensional graph cosmohedra for the path graph, cycle graph, and complete graph on four vertices. In the case of the path graph and complete graph we recover the cosmohedron and the permutoassociahedron respectively. }
    \label{fig:cosmo_exam}
\end{figure}

First, for each tubing $\tau$ on $G$ we introduce a new variable $Y_\tau$ and a cut parameter $\epsilon_\tau$ and set
\begin{equation}
Y_\tau=\sum_{t\in \tau}X_t-\epsilon_\tau\,.
\end{equation}
As in the case of path graphs in \cite{Arkani-Hamed:2024jbp}, the $\epsilon$ parameters must satisfy certain conditions to correctly cut out the graph cosmohedron. In particular, we demand that
\begin{equation}
\epsilon_\tau+\epsilon_{\tau'}= \epsilon_{\tau\cup \tau'}+\epsilon_{\tau\cap \tau'},
\label{eq:submod}
\end{equation}
for every pair of compatible tubings $\tau$ and $\tau'$ such that
\begin{itemize}
\item the regions in $R(\tau)\setminus R(\tau\cap \tau')$ are all nested inside a single region $r_1$ of $R(\tau\cap \tau')$, 
\item the regions in $R(\tau')\setminus R(\tau\cap \tau')$ are all nested inside a single region $r_2$ of $R(\tau\cap \tau')$,
\item and $r_1\neq r_2$.
\end{itemize} 
An example of a configuration satisfying \eqref{eq:submod} for the graph $P_4$ is given by
\begin{align}
\epsilon_{\includegraphics[scale=0.7]{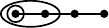}}+\epsilon_{\includegraphics[scale=0.7]{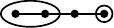}}=\epsilon_{\includegraphics[scale=0.7]{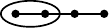}}+\epsilon_{\includegraphics[scale=0.7]{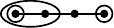}},
\end{align}
since
\begin{align}
\raisebox{-0.8ex}{\includegraphics[scale=1]{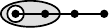}} \ \sqcup \  \raisebox{-0.35ex}{\includegraphics[scale=1]{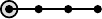}} = \raisebox{-0.8ex}{\includegraphics[scale=1]{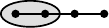}} \  \neq \  \raisebox{-1.3ex}{\includegraphics[scale=1]{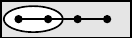}}=\raisebox{-1.3ex}{\includegraphics[scale=1]{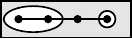}} \ \sqcup \  \raisebox{-0.35ex}{\includegraphics[scale=1]{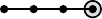}}.
\end{align}
Additionally, for all other pairs of compatible tubings $\tau$ and $\tau'$ we require 
\begin{equation}
\epsilon_\tau+\epsilon_{\tau'}< \epsilon_{\tau\cup \tau'}+\epsilon_{\tau\cap \tau'}.
\label{eq:submod2}
\end{equation}
An example of a configuration where the inequality is not saturated, again for the graph $P_4$, is given by
\begin{align}
\epsilon_{\includegraphics[scale=0.7]{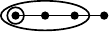}}+\epsilon_{\includegraphics[scale=0.7]{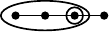}}<\epsilon_{\includegraphics[scale=0.7]{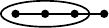}}+\epsilon_{\includegraphics[scale=0.7]{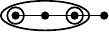}},
\end{align}
since we have the following
\begin{align}
\raisebox{-0.98ex}{\includegraphics[scale=1]{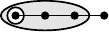}} \ \sqcup \  \raisebox{-0.35ex}{\includegraphics[scale=1]{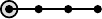}}=\raisebox{-0.98ex}{\includegraphics[scale=1]{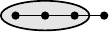}}=\raisebox{-0.98ex}{\includegraphics[scale=1]{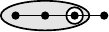}} \ \sqcup \  \raisebox{-0.35ex}{\includegraphics[scale=1]{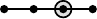}}.
\end{align}
The set of equalities in \eqref{eq:submod} are trivially satisfied if the cut parameters $\epsilon_\tau$ are written as a sum over {\it regions} of the tubing $\tau$ as 
\begin{equation}
\epsilon_\tau=\sum_{r\in R(\tau)} \delta_r,
\end{equation}
where we introduced new parameters $\delta_r$ for each region of $G$.
Finally, the inequalities \eqref{eq:submod2} are satisfied whenever $\delta_r$ is a convex function of $|V_r|$ that vanishes on the region specified by the root: $r=\{V_G\}$. A particular example of such a function is
\begin{equation}
\delta_r= \delta \sum_{I\subset \overline{V_r}}p_I,
\end{equation}
where $\overline{V_r}$ is the complement of the set of vertices of $r$ in $V_G$, $p_I>0$ for all $I\subset V_G$ with $|I|>1$, and $0<\delta\ll 1 $ is a small positive parameter. We also set $p_{\{v\}}=0$ for all $v\in V_G$. With this notation, we define the graph cosmohedron $\mathcal{C}_G$ as the intersection of the collection of half spaces defined by $Y_\tau\geq 0$ for all tubings on $G$:
\begin{align}\label{eq_ABHY_cosmo}
\boxed{\mathcal{C}_G=\{ (X_{\{1 \}},\ldots,X_{\{|G| \}}) \in \mathbb{R}^{|G|} :   (\forall_{\tau\in\Gamma_G} \,Y_\tau\geq0)\wedge (X_{V_G}=0)\},}
\end{align}
restricted to the $X_{V_G}=0$ hyperplane.
Explicit examples of the embedding of the graph cosmohedra for the path, cycle and complete graphs on four vertices are displayed in Fig.~\ref{fig:cosmo_exam}.

\subsection{Examples of Graph Cosmohedra}
\begin{figure}[t]
    \centering
    \includegraphics[width=0.6\textwidth]{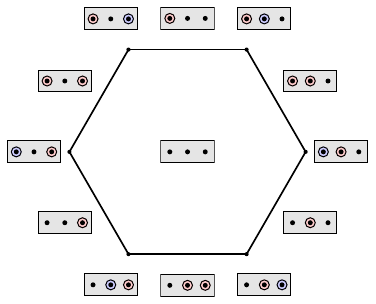}
    \caption{The graph cosmohedron for the empty graph on three vertices which reproduces the two dimensional permutohedron. }
    \label{fig:cosmo_E3_exam}
\end{figure}

Let us demonstrate the graph cosmohedron inequalities in some simple examples. We begin with the empty graph on three vertices $N_3$. In this case the inequalities of the graph cosmohedron are given by a modification of those of \eqref{eq:v3_empty} 
\begin{align}
& X_1  \geq \delta \,p_{\{2,3\}}, \quad X_2 \geq \delta \,p_{\{1,3\}}, \quad X_3 \geq \delta \,p_{\{1,2\}},
\end{align}
together with the additional inequalities 
\begin{align}
& X_1 +X_2 \geq \delta(p_{\{1,2\}}+p_{\{1,3\}}+p_{\{2,3\}}), \notag \\
& X_1 +X_3 \geq \delta(p_{\{1,2\}}+p_{\{1,3\}}+p_{\{2,3\}}), \notag \\
& X_2 +X_3 \geq \delta(p_{\{1,2\}}+p_{\{1,3\}}+p_{\{2,3\}}).
\end{align}
Then the cosmohedron $\mathcal{C}_{N_3}$ coincides with the two-dimensional permutohedron displayed in Fig.~\ref{fig:cosmo_E3_exam}. More generally, the graph cosmohedron for the empty graph on $(d+1)$ vertices reproduces the $d$-dimensional permutohedron. An explicit embedding of the three-dimensional example is displayed in Fig.~\ref{fig:cosmo_exam}.

Next consider the path graph $P_3$ on three vertices. The inequalities carving out the cosmohedron in this case are given by the modified version of \eqref{eq:v3_path}, namely
\begin{align}
& X_1  \geq \delta\, p_{\{2,3\}}, \quad X_2 \geq \delta \,p_{\{1,3\}}, \quad X_3 \geq \delta \,p_{\{1,2\}}, \notag \\
&X_1+X_2 \geq c_{\{1,2\}}+ \delta \,p_{\{1,2\}}, \quad X_2+X_3 \geq c_{\{2,3\}}+\delta \, p_{\{2,3\}}.
\end{align}
In addition, we have five new inequalities
\begin{align}
&X_1+X_3 \geq  \delta(p_{\{1,2\}}+p_{\{1,3\}}+p_{\{2,3\}}), \notag \\
&2X_1+X_2 \geq  c_{\{1,2\}}+\delta(p_{\{1,2\}}+p_{\{1,3\}}+p_{\{2,3\}}), \notag \\
&X_1+2X_2 \geq  c_{\{1,2\}}+\delta(p_{\{1,2\}}+p_{\{1,3\}}+p_{\{2,3\}}), \notag \\
&2X_2+X_3 \geq  c_{\{2,3\}}+\delta(p_{\{1,2\}}+p_{\{1,3\}}+p_{\{2,3\}}), \notag \\
&X_2+2X_3 \geq  c_{\{2,3\}}+\delta(p_{\{1,2\}}+p_{\{1,3\}}+p_{\{2,3\}}).
\end{align}
As the result, the graph cosmohedron $\mathcal{C}_{P_3}$ is the decagon displayed in Fig.~\ref{fig:cosmo_P3_exam}. For arbitrary path graphs, one recovers the family of cosmohedra defined in  \cite{Arkani-Hamed:2024jbp}. A three dimensional example is shown in Fig.~\ref{fig:cosmo_exam}. 

\begin{figure}[t]
    \centering
    \includegraphics[width=0.6\textwidth]{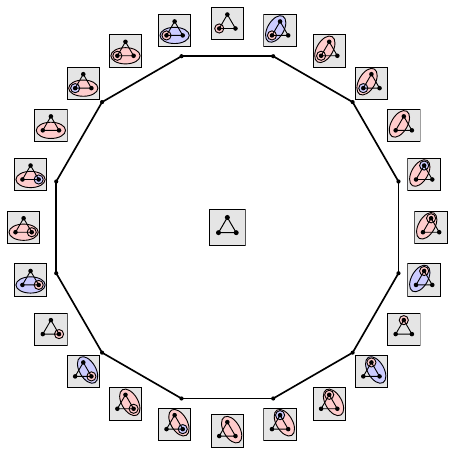}
    \caption{The graph cosmohedron for the complete graph on three vertices which reproduces the two dimensional permutoassociahedron.}
    \label{fig:cosmo_K3_exam}
\end{figure}

Finally, we consider the complete graph on $(d+1)$ vertices. In this case the graph cosmohedron produces the family of polytopes known as permutoassociahedra introduced in \cite{kapranov1993permutoassociahedron}. The two-dimensional permutoassociahedron with boundaries labelled by regional tubings is displayed in Fig.~\ref{fig:cosmo_K3_exam}. An explicit embedding of the three-dimensional example is displayed in Fig.~\ref{fig:cosmo_exam}. We provide a list of combinatorial data associated to graph cosmohedra for various graphs with two, three and four vertices in Table \ref{tab:cosmo}.

\begin{table}
\begin{center}
\begin{tabular}{c|c|c|c|c}
Graph $G$&dimension& codim-1&codim-2&codim-3\\
\hline
$\tikz[baseline,yshift=0.5ex] \node[anchor=base] at (0,0) {\begin{tikzpicture}
        \coordinate (A) at (0,0);
        \coordinate (B) at (1/2,0);
        \fill[black] (A) circle (2pt);
        \fill[black] (B) circle (2pt);
    \end{tikzpicture}};$ &1&2&&\\
\hline
$\tikz[baseline,yshift=0.5ex] \node[anchor=base] at (0,0) {\begin{tikzpicture}
        \coordinate (A) at (0,0);
        \coordinate (B) at (1/2,0);
        \draw[thick] (A) -- (B);
        \fill[black] (A) circle (2pt);
        \fill[black] (B) circle (2pt);
    \end{tikzpicture}};$ &1&2&&\\
\hline
\hline
$\tikz[baseline,yshift=0.5ex] \node[anchor=base] at (0,0) {\begin{tikzpicture}
        \coordinate (A) at (0,0);
        \coordinate (B) at (1/2,0);
        \coordinate (C) at (1,0);
        \fill[black] (A) circle (2pt);
        \fill[black] (B) circle (2pt);
        \fill[black] (C) circle (2pt);
    \end{tikzpicture}};$ &2&6&6&\\
    \hline
$\tikz[baseline,yshift=0.5ex] \node[anchor=base] at (0,0) {\begin{tikzpicture}
        \coordinate (A) at (0,0);
        \coordinate (B) at (1/2,0);
        \coordinate (C) at (1,0);
        \draw[thick] (A) -- (B) ;
        \fill[black] (A) circle (2pt);
        \fill[black] (B) circle (2pt);
        \fill[black] (C) circle (2pt);
    \end{tikzpicture}};$ &2&8&8&\\
    \hline
$\tikz[baseline,yshift=0.5ex] \node[anchor=base] at (0,0) {\begin{tikzpicture}
        \coordinate (A) at (0,0);
        \coordinate (B) at (1/2,0);
        \coordinate (C) at (1,0);
        \draw[thick] (A) -- (B) -- (C);
        \fill[black] (A) circle (2pt);
        \fill[black] (B) circle (2pt);
        \fill[black] (C) circle (2pt);
    \end{tikzpicture}};$ &2&10&10&\\
        \hline
$\tikz[baseline,yshift=0.5ex] \node[anchor=base] at (0,0) {\begin{tikzpicture}
        \coordinate (A) at (0,0);
        \coordinate (B) at (1/2,0);
        \coordinate (C) at (1/4,1/3);
        \draw[thick] (A) -- (B) -- (C) -- (A);
        \fill[black] (A) circle (2pt);
        \fill[black] (B) circle (2pt);
        \fill[black] (C) circle (2pt);
    \end{tikzpicture}};$ &2&12&12&\\
    \hline
    \hline
$\tikz[baseline,yshift=0.5ex] \node[anchor=base] at (0,0) {\begin{tikzpicture}
        \coordinate (A) at (0,0);
        \coordinate (B) at (1/2,0);
        \coordinate (C) at (1,0);
        \coordinate (D) at (3/2,0);
        \draw[thick] (A) -- (B) -- (C) -- (D);
        \fill[black] (A) circle (2pt);
        \fill[black] (B) circle (2pt);
        \fill[black] (C) circle (2pt);
        \fill[black] (D) circle (2pt);
    \end{tikzpicture}};$ &3&44&114&72\\
\hline
$\tikz[baseline,yshift=0.5ex] \node[anchor=base] at (0,0) {\begin{tikzpicture}
       \coordinate (A) at (0,0);
        \coordinate (B) at (1/4,1/4);
        \coordinate (C) at (-1/4,1/4);
        \coordinate (D) at (0,-1/3);
        \draw[thick] (B) -- (A) -- (C);
                \draw[thick] (D) -- (A);
        \fill[black] (A) circle (2pt);
        \fill[black] (B) circle (2pt);
        \fill[black] (C) circle (2pt);
        \fill[black] (D) circle (2pt);
    \end{tikzpicture}};$ &3&50&132&84\\
\hline
$\tikz[baseline,yshift=0.5ex] \node[anchor=base] at (0,0) {\begin{tikzpicture}
        \coordinate (A) at (0,0);
        \coordinate (B) at (1/2,0);
        \coordinate (C) at (1/2,1/2);
        \coordinate (D) at (0,1/2);
        \draw[thick] (A) -- (B) -- (C) -- (D)--(A);
        \fill[black] (A) circle (2pt);
        \fill[black] (B) circle (2pt);
        \fill[black] (C) circle (2pt);
        \fill[black] (D) circle (2pt);
    \end{tikzpicture}};$ &3&62&164&104\\
    \hline
$\tikz[baseline,yshift=0.5ex] \node[anchor=base] at (0,0) {\begin{tikzpicture}
        \coordinate (A) at (0,0);
        \coordinate (B) at (1/2,0);
        \coordinate (C) at (1/2,1/2);
        \coordinate (D) at (0,1/2);
        \draw[thick] (A) -- (B) -- (C) -- (D)--(A);
        \draw[thick] (A) -- (C);
        \draw[thick]  (B) --(D);
        \fill[black] (A) circle (2pt);
        \fill[black] (B) circle (2pt);
        \fill[black] (C) circle (2pt);
        \fill[black] (D) circle (2pt);
    \end{tikzpicture}};$ &3&74&192&120\\
\end{tabular}
\end{center}
\caption{The number of boundaries of various codimension of the cosmohedron on graph $G$.}
\label{tab:cosmo}
\end{table}
\subsection{Canonical Forms for Graph Cosmohedra}
\label{sec:can_form}
Since a generic graph cosmohedron is not a simple polytope, its canonical form cannot be obtained from the formula \eqref{eq:can_form_assoc} that we used for the graph associahedra. However, one can use its modification, inspired by the recent results in \cite{Brown:2025jjg}, where the canonical form of any convex polytope can be written as a linear combination of terms coming from all vertices. For a $d$-dimensional graph cosmohedron, the formula is 
\begin{equation}
\Omega(\mathcal{C}_G)=\sum_{\rho\in \Phi_G^{\text{max}}}\sum_{I\in \binom{[F_\rho]}{d}} \alpha_{\rho,I}\bigwedge_{\tilde{\rho}\in I} \dd \log (Y_{\tilde{\rho}}),
\end{equation}
where the first sum is over all vertices of the cosmohedron, the second sum is over all $d$-element subsets of faces meeting at the vertex specified by $\rho$. The coefficients $\alpha_{\rho,I}$ can be fixed on the case-by-case basis by demanding that the form $\Omega_{\mathcal{C}_G}$ is projectively invariant.

After pulling back to the appropriate subspace and stripping off the measure, the last step needed in order to recover the cosmological amplitube is to make the following substitution for the $Y$ variables 
\begin{align}
Y_{\rho} =\sum_{t \in B(\rho)} \frac{1}{\mathcal{R}^{(1)}_{t,\rho} \mathcal{R}^{(2)}_{t,\rho}},
\label{eq:YtoRR}
\end{align}
where $\mathcal{R}^{(i)}_{t,\rho}$, $i=1,2$, correspond to the unique pair of regions in $\rho$ which share the tube $t$ as a border. Having made the replacements \eqref{eq:YtoRR} and keeping only terms with simple poles we arrive at the expression for the corresponding cosmological amplitube $\Psi_G$.

We demonstrate the above procedure with an example. The first interesting case where the graph cosmohedron is not simple is for the path graph $P_4$ on four vertices. In this case the graph cosmohedron $\mathcal{C}_{P_4}$ has $12$ non-simple vertices, an example of which is given by
\begin{align}
\rho_v= \tikz[baseline,yshift=0.5ex] \node[anchor=base] at (0,0) {\begin{tikzpicture}
        \coordinate (A) at (0,0);
        \coordinate (B) at (1/2,0);
        \coordinate (C) at (1,0);
        \coordinate (D) at (3/2,0);
        \fillroundrectangle{0}{0.8}{0}{2.2}{0.62}{0};
         \roundrectangle{0}{0.8}{0}{2.2}{0.62}{0};
        \filldraw[thick, red!20] (0.25,0) ellipse (0.5cm and 0.22cm);
        \draw[thick, black] (0.25,0) ellipse (0.5cm and 0.22cm);
        \filldraw[blue!25] (A) circle (4pt);
        \draw[black,thick] (A) circle (4pt);
        \filldraw[blue!25] (D) circle (4pt);
        \draw[black,thick] (D) circle (4pt);
        \draw[thick] (A) -- (B) -- (C) -- (D);
        \fill[black] (A) circle (2pt);
        \fill[black] (B) circle (2pt);
        \fill[black] (C) circle (2pt);
        \fill[black] (D) circle (2pt);
    \end{tikzpicture}};.
\end{align} 
This vertex is defined by the intersection of the following four facets
\begin{align}
\rho_v^{(1)}= \tikz[baseline,yshift=0.5ex] \node[anchor=base] at (0,0) {\begin{tikzpicture}[scale=0.8]
        \coordinate (A) at (0,0);
        \coordinate (B) at (1/2,0);
        \coordinate (C) at (1,0);
        \coordinate (D) at (3/2,0);
        \fillroundrectangle{0}{0.8}{0}{2.2}{0.62}{0};
         \roundrectangle{0}{0.8}{0}{2.2}{0.62}{0};
        \filldraw[thick, red!20] (0.25,0) ellipse (0.5cm and 0.22cm);
        \draw[thick, black] (0.25,0) ellipse (0.5cm and 0.22cm);
        \draw[black,thick] (A) circle (4pt);
        \filldraw[red!20] (D) circle (4pt);
        \draw[black,thick] (D) circle (4pt);
        \draw[thick] (A) -- (B) -- (C) -- (D);
        \fill[black] (A) circle (2pt);
        \fill[black] (B) circle (2pt);
        \fill[black] (C) circle (2pt);
        \fill[black] (D) circle (2pt);
    \end{tikzpicture}};, \quad
    \label{eq:four_facets}
     \rho_v^{(2)}=\tikz[baseline,yshift=0.5ex] \node[anchor=base] at (0,0) {\begin{tikzpicture}[scale=0.8]
        \coordinate (A) at (0,0);
        \coordinate (B) at (1/2,0);
        \coordinate (C) at (1,0);
        \coordinate (D) at (3/2,0);
        \fillroundrectangle{0}{0.8}{0}{2.2}{0.62}{0};
         \roundrectangle{0}{0.8}{0}{2.2}{0.62}{0};
         \filldraw[thick, red!20] (0.25,0) ellipse (0.5cm and 0.22cm);
        \draw[thick, black] (0.25,0) ellipse (0.5cm and 0.22cm);
        \filldraw[red!20] (D) circle (4pt);
        \draw[black,thick] (D) circle (4pt);
        \draw[thick] (A) -- (B) -- (C) -- (D);
        \fill[black] (A) circle (2pt);
        \fill[black] (B) circle (2pt);
        \fill[black] (C) circle (2pt);
        \fill[black] (D) circle (2pt);
    \end{tikzpicture}};,
      \quad 
      \rho_v^{(3)}=\tikz[baseline,yshift=0.5ex] \node[anchor=base] at (0,0) {\begin{tikzpicture}[scale=0.8]
        \coordinate (A) at (0,0);
        \coordinate (B) at (1/2,0);
        \coordinate (C) at (1,0);
        \coordinate (D) at (3/2,0);
        \fillroundrectangle{0}{0.8}{0}{2.2}{0.62}{0};
         \roundrectangle{0}{0.8}{0}{2.2}{0.62}{0};
        \filldraw[thick, red!20] (0.25,0) ellipse (0.5cm and 0.22cm);
        \draw[thick, black] (0.25,0) ellipse (0.5cm and 0.22cm);
        \draw[thick] (A) -- (B) -- (C) -- (D);
        \fill[black] (A) circle (2pt);
        \fill[black] (B) circle (2pt);
        \fill[black] (C) circle (2pt);
        \fill[black] (D) circle (2pt);
    \end{tikzpicture}};,
    \rho_v^{(4)}=\tikz[baseline,yshift=0.5ex] \node[anchor=base] at (0,0) {\begin{tikzpicture}[scale=0.8]
        \coordinate (A) at (0,0);
        \coordinate (B) at (1/2,0);
        \coordinate (C) at (1,0);
        \coordinate (D) at (3/2,0);
        \fillroundrectangle{0}{0.8}{0}{2.2}{0.62}{0};
         \roundrectangle{0}{0.8}{0}{2.2}{0.62}{0};
        \filldraw[thick, red!20] (0.25,0) ellipse (0.5cm and 0.22cm);
        \draw[thick, black] (0.25,0) ellipse (0.5cm and 0.22cm);
        \draw[black,thick] (A) circle (4pt);
        \draw[thick] (A) -- (B) -- (C) -- (D);
        \fill[black] (A) circle (2pt);
        \fill[black] (B) circle (2pt);
        \fill[black] (C) circle (2pt);
        \fill[black] (D) circle (2pt);
    \end{tikzpicture}};,
\end{align}
and the contribution of this vertex to the canonical form $\Omega(\mathcal{C}_{P_4})$ is
\begin{equation}
\Omega(\mathcal{C}_{P_4})=\dd \log \frac{Y_{\tikz[baseline,yshift=0.5ex] \node[anchor=base] at (0,0) {\begin{tikzpicture}[scale=0.5]
        \coordinate (A) at (0,0);
        \coordinate (B) at (1/2,0);
        \coordinate (C) at (1,0);
        \coordinate (D) at (3/2,0);
        \fillroundrectangle{0}{0.8}{0}{2.2}{0.62}{0};
         \roundrectangle{0}{0.8}{0}{2.2}{0.62}{0};
        \filldraw[thick, red!20] (0.25,0) ellipse (0.5cm and 0.22cm);
        \draw[thick, black] (0.25,0) ellipse (0.5cm and 0.22cm);
        \draw[black,thick] (A) circle (4pt);
        \filldraw[red!20] (D) circle (4pt);
        \draw[black,thick] (D) circle (4pt);
        \draw[thick] (A) -- (B) -- (C) -- (D);
        \fill[black] (A) circle (2pt);
        \fill[black] (B) circle (2pt);
        \fill[black] (C) circle (2pt);
        \fill[black] (D) circle (2pt);
    \end{tikzpicture}};}}{Y_{\tikz[baseline,yshift=0.5ex] \node[anchor=base] at (0,0) {\begin{tikzpicture}[scale=0.5]
        \coordinate (A) at (0,0);
        \coordinate (B) at (1/2,0);
        \coordinate (C) at (1,0);
        \coordinate (D) at (3/2,0);
        \fillroundrectangle{0}{0.8}{0}{2.2}{0.62}{0};
         \roundrectangle{0}{0.8}{0}{2.2}{0.62}{0};
        \filldraw[thick, red!20] (0.25,0) ellipse (0.5cm and 0.22cm);
        \draw[thick, black] (0.25,0) ellipse (0.5cm and 0.22cm);
        \draw[thick] (A) -- (B) -- (C) -- (D);
        \fill[black] (A) circle (2pt);
        \fill[black] (B) circle (2pt);
        \fill[black] (C) circle (2pt);
        \fill[black] (D) circle (2pt);
    \end{tikzpicture}};}}\wedge \dd \log Y_{\tikz[baseline,yshift=0.5ex] \node[anchor=base] at (0,0) {\begin{tikzpicture}[scale=0.5]
        \coordinate (A) at (0,0);
        \coordinate (B) at (1/2,0);
        \coordinate (C) at (1,0);
        \coordinate (D) at (3/2,0);
        \fillroundrectangle{0}{0.8}{0}{2.2}{0.62}{0};
         \roundrectangle{0}{0.8}{0}{2.2}{0.62}{0};
         \filldraw[thick, red!20] (0.25,0) ellipse (0.5cm and 0.22cm);
        \draw[thick, black] (0.25,0) ellipse (0.5cm and 0.22cm);
        \filldraw[red!20] (D) circle (4pt);
        \draw[black,thick] (D) circle (4pt);
        \draw[thick] (A) -- (B) -- (C) -- (D);
        \fill[black] (A) circle (2pt);
        \fill[black] (B) circle (2pt);
        \fill[black] (C) circle (2pt);
        \fill[black] (D) circle (2pt);
    \end{tikzpicture}};}\wedge \dd \log Y_{\tikz[baseline,yshift=0.5ex] \node[anchor=base] at (0,0) {\begin{tikzpicture}[scale=0.5]
        \coordinate (A) at (0,0);
        \coordinate (B) at (1/2,0);
        \coordinate (C) at (1,0);
        \coordinate (D) at (3/2,0);
        \fillroundrectangle{0}{0.8}{0}{2.2}{0.62}{0};
         \roundrectangle{0}{0.8}{0}{2.2}{0.62}{0};
        \filldraw[thick, red!20] (0.25,0) ellipse (0.5cm and 0.22cm);
        \draw[thick, black] (0.25,0) ellipse (0.5cm and 0.22cm);
        \draw[black,thick] (A) circle (4pt);
        \draw[thick] (A) -- (B) -- (C) -- (D);
        \fill[black] (A) circle (2pt);
        \fill[black] (B) circle (2pt);
        \fill[black] (C) circle (2pt);
        \fill[black] (D) circle (2pt);
    \end{tikzpicture}};}+\ldots\,.
    \label{eq:CP4_vertex}
\end{equation}
After pulling back to the appropriate subspace and stripping off the measure, we make the following substitutions
\begin{align}
\frac{1}{Y_{\tikz[baseline,yshift=0.5ex] \node[anchor=base] at (0,0) {\begin{tikzpicture}[scale=0.5]
        \coordinate (A) at (0,0);
        \coordinate (B) at (1/2,0);
        \coordinate (C) at (1,0);
        \coordinate (D) at (3/2,0);
        \fillroundrectangle{0}{0.8}{0}{2.2}{0.62}{0};
         \roundrectangle{0}{0.8}{0}{2.2}{0.62}{0};
        \filldraw[thick, red!20] (0.25,0) ellipse (0.5cm and 0.22cm);
        \draw[thick, black] (0.25,0) ellipse (0.5cm and 0.22cm);
        \draw[black,thick] (A) circle (4pt);
        \filldraw[red!20] (D) circle (4pt);
        \draw[black,thick] (D) circle (4pt);
        \draw[thick] (A) -- (B) -- (C) -- (D);
        \fill[black] (A) circle (2pt);
        \fill[black] (B) circle (2pt);
        \fill[black] (C) circle (2pt);
        \fill[black] (D) circle (2pt);
    \end{tikzpicture}};}} &= \frac{1}{\mathcal{R}_{\tikz[baseline,yshift=0.5ex] \node[anchor=base] at (0,0) {\begin{tikzpicture}[scale=0.5]
        \coordinate (A) at (0,0);
        \coordinate (B) at (1/2,0);
        \coordinate (C) at (1,0);
        \coordinate (D) at (3/2,0);
         \draw[thick] (A) -- (B) -- (C) -- (D);
        \filldraw[gray!20] (A) circle (4pt);
        \draw[black,thick] (A) circle (4pt);
        \fill[black] (A) circle (2pt);
        \fill[black] (B) circle (2pt);
        \fill[black] (C) circle (2pt);
        \fill[black] (D) circle (2pt);
    \end{tikzpicture}};}\mathcal{R}_{\tikz[baseline,yshift=0.5ex] \node[anchor=base] at (0,0) {\begin{tikzpicture}[scale=0.5]
        \coordinate (A) at (0,0);
        \coordinate (B) at (1/2,0);
        \coordinate (C) at (1,0);
        \coordinate (D) at (3/2,0);
        \filldraw[thick, gray!20] (0.25,0) ellipse (0.5cm and 0.22cm);
        \draw[thick, black] (0.25,0) ellipse (0.5cm and 0.22cm);
         \draw[thick] (A) -- (B) -- (C) -- (D);
        \filldraw[white] (A) circle (4pt);
        \draw[black,thick] (A) circle (4pt);
        \fill[black] (A) circle (2pt);
        \fill[black] (B) circle (2pt);
        \fill[black] (C) circle (2pt);
        \fill[black] (D) circle (2pt);
    \end{tikzpicture}};}}+\frac{1}{\mathcal{R}_{\tikz[baseline,yshift=0.5ex] \node[anchor=base] at (0,0) {\begin{tikzpicture}[scale=0.5]
        \coordinate (A) at (0,0);
        \coordinate (B) at (1/2,0);
        \coordinate (C) at (1,0);
        \coordinate (D) at (3/2,0);
        \filldraw[thick, gray!20] (0.25,0) ellipse (0.5cm and 0.22cm);
        \draw[thick, black] (0.25,0) ellipse (0.5cm and 0.22cm);
         \draw[thick] (A) -- (B) -- (C) -- (D);
        \filldraw[white] (A) circle (4pt);
        \draw[black,thick] (A) circle (4pt);
        \fill[black] (A) circle (2pt);
        \fill[black] (B) circle (2pt);
        \fill[black] (C) circle (2pt);
        \fill[black] (D) circle (2pt);
    \end{tikzpicture}};}\mathcal{R}_{\tikz[baseline,yshift=0.5ex] \node[anchor=base] at (0,0) {\begin{tikzpicture}[scale=0.5]
        \coordinate (A) at (0,0);
        \coordinate (B) at (1/2,0);
        \coordinate (C) at (1,0);
        \coordinate (D) at (3/2,0);
        \fillroundrectangle{0}{0.8}{0}{2.2}{0.62}{0};
         \roundrectangle{0}{0.8}{0}{2.2}{0.62}{0};
        \filldraw[thick, white] (.25,0) ellipse (0.45cm and 0.2cm);
        \draw[thick, black] (.25,0) ellipse (0.45cm and 0.2cm);
         \draw[thick] (A) -- (B) -- (C) -- (D);
        \filldraw[white] (D) circle (4pt);
        \draw[black,thick] (D) circle (4pt);
        \fill[black] (A) circle (2pt);
        \fill[black] (B) circle (2pt);
        \fill[black] (C) circle (2pt);
        \fill[black] (D) circle (2pt);
    \end{tikzpicture}};}}+\frac{1}{\mathcal{R}_{ \tikz[baseline,yshift=0.5ex] \node[anchor=base] at (0,0) {\begin{tikzpicture}[scale=0.5]
        \coordinate (A) at (0,0);
        \coordinate (B) at (1/2,0);
        \coordinate (C) at (1,0);
        \coordinate (D) at (3/2,0);
         \draw[thick] (A) -- (B) -- (C) -- (D);
        \filldraw[gray!20] (D) circle (4pt);
        \draw[black,thick] (D) circle (4pt);
        \fill[black] (A) circle (2pt);
        \fill[black] (B) circle (2pt);
        \fill[black] (C) circle (2pt);
        \fill[black] (D) circle (2pt);
    \end{tikzpicture}};}\mathcal{R}_{\tikz[baseline,yshift=0.5ex] \node[anchor=base] at (0,0) {\begin{tikzpicture}[scale=0.5]
        \coordinate (A) at (0,0);
        \coordinate (B) at (1/2,0);
        \coordinate (C) at (1,0);
        \coordinate (D) at (3/2,0);
        \fillroundrectangle{0}{0.8}{0}{2.2}{0.62}{0};
         \roundrectangle{0}{0.8}{0}{2.2}{0.62}{0};
        \filldraw[thick, white] (.25,0) ellipse (0.45cm and 0.2cm);
        \draw[thick, black] (.25,0) ellipse (0.45cm and 0.2cm);
         \draw[thick] (A) -- (B) -- (C) -- (D);
        \filldraw[white] (D) circle (4pt);
        \draw[black,thick] (D) circle (4pt);
        \fill[black] (A) circle (2pt);
        \fill[black] (B) circle (2pt);
        \fill[black] (C) circle (2pt);
        \fill[black] (D) circle (2pt);
    \end{tikzpicture}};}}, \notag \\
\frac{1}{Y_{\tikz[baseline,yshift=0.5ex] \node[anchor=base] at (0,0) {\begin{tikzpicture}[scale=0.5]
        \coordinate (A) at (0,0);
        \coordinate (B) at (1/2,0);
        \coordinate (C) at (1,0);
        \coordinate (D) at (3/2,0);
        \fillroundrectangle{0}{0.8}{0}{2.2}{0.62}{0};
         \roundrectangle{0}{0.8}{0}{2.2}{0.62}{0};
         \filldraw[thick, red!20] (0.25,0) ellipse (0.5cm and 0.22cm);
        \draw[thick, black] (0.25,0) ellipse (0.5cm and 0.22cm);
        \filldraw[red!20] (D) circle (4pt);
        \draw[black,thick] (D) circle (4pt);
        \draw[thick] (A) -- (B) -- (C) -- (D);
        \fill[black] (A) circle (2pt);
        \fill[black] (B) circle (2pt);
        \fill[black] (C) circle (2pt);
        \fill[black] (D) circle (2pt);
    \end{tikzpicture}};}} &= \frac{1}{\mathcal{R}_{\tikz[baseline,yshift=0.5ex] \node[anchor=base] at (0,0) {\begin{tikzpicture}[scale=0.5]
        \coordinate (A) at (0,0);
        \coordinate (B) at (1/2,0);
        \coordinate (C) at (1,0);
        \coordinate (D) at (3/2,0);
        \filldraw[thick, gray!20] (0.25,0) ellipse (0.4cm and 0.17cm);
        \draw[thick, black] (0.25,0) ellipse (0.4cm and 0.17cm);
         \draw[thick] (A) -- (B) -- (C) -- (D);
        \fill[black] (A) circle (2pt);
        \fill[black] (B) circle (2pt);
        \fill[black] (C) circle (2pt);
        \fill[black] (D) circle (2pt);
    \end{tikzpicture}};}\mathcal{R}_{\tikz[baseline,yshift=0.5ex] \node[anchor=base] at (0,0) {\begin{tikzpicture}[scale=0.5]
        \coordinate (A) at (0,0);
        \coordinate (B) at (1/2,0);
        \coordinate (C) at (1,0);
        \coordinate (D) at (3/2,0);
        \fillroundrectangle{0}{0.8}{0}{2.2}{0.62}{0};
         \roundrectangle{0}{0.8}{0}{2.2}{0.62}{0};
        \filldraw[thick, white] (.25,0) ellipse (0.45cm and 0.2cm);
        \draw[thick, black] (.25,0) ellipse (0.45cm and 0.2cm);
         \draw[thick] (A) -- (B) -- (C) -- (D);
        \filldraw[white] (D) circle (4pt);
        \draw[black,thick] (D) circle (4pt);
        \fill[black] (A) circle (2pt);
        \fill[black] (B) circle (2pt);
        \fill[black] (C) circle (2pt);
        \fill[black] (D) circle (2pt);
    \end{tikzpicture}};}}+\frac{1}{\mathcal{R}_{ \tikz[baseline,yshift=0.5ex] \node[anchor=base] at (0,0) {\begin{tikzpicture}[scale=0.5]
        \coordinate (A) at (0,0);
        \coordinate (B) at (1/2,0);
        \coordinate (C) at (1,0);
        \coordinate (D) at (3/2,0);
         \draw[thick] (A) -- (B) -- (C) -- (D);
        \filldraw[gray!20] (D) circle (4pt);
        \draw[black,thick] (D) circle (4pt);
        \fill[black] (A) circle (2pt);
        \fill[black] (B) circle (2pt);
        \fill[black] (C) circle (2pt);
        \fill[black] (D) circle (2pt);
    \end{tikzpicture}};} \mathcal{R}_{\tikz[baseline,yshift=0.5ex] \node[anchor=base] at (0,0) {\begin{tikzpicture}[scale=0.5]
        \coordinate (A) at (0,0);
        \coordinate (B) at (1/2,0);
        \coordinate (C) at (1,0);
        \coordinate (D) at (3/2,0);
        \fillroundrectangle{0}{0.8}{0}{2.2}{0.62}{0};
         \roundrectangle{0}{0.8}{0}{2.2}{0.62}{0};
        \filldraw[thick, white] (.25,0) ellipse (0.45cm and 0.2cm);
        \draw[thick, black] (.25,0) ellipse (0.45cm and 0.2cm);
         \draw[thick] (A) -- (B) -- (C) -- (D);
        \filldraw[white] (D) circle (4pt);
        \draw[black,thick] (D) circle (4pt);
        \fill[black] (A) circle (2pt);
        \fill[black] (B) circle (2pt);
        \fill[black] (C) circle (2pt);
        \fill[black] (D) circle (2pt);
    \end{tikzpicture}};}}, \notag\\
\frac{1}{Y_{\tikz[baseline,yshift=0.5ex] \node[anchor=base] at (0,0) {\begin{tikzpicture}[scale=0.5]
        \coordinate (A) at (0,0);
        \coordinate (B) at (1/2,0);
        \coordinate (C) at (1,0);
        \coordinate (D) at (3/2,0);
        \fillroundrectangle{0}{0.8}{0}{2.2}{0.62}{0};
         \roundrectangle{0}{0.8}{0}{2.2}{0.62}{0};
        \filldraw[thick, red!20] (0.25,0) ellipse (0.5cm and 0.22cm);
        \draw[thick, black] (0.25,0) ellipse (0.5cm and 0.22cm);
        \draw[thick] (A) -- (B) -- (C) -- (D);
        \fill[black] (A) circle (2pt);
        \fill[black] (B) circle (2pt);
        \fill[black] (C) circle (2pt);
        \fill[black] (D) circle (2pt);
    \end{tikzpicture}};}} &= \frac{1}{\mathcal{R}_{\tikz[baseline,yshift=0.5ex] \node[anchor=base] at (0,0) {\begin{tikzpicture}[scale=0.5]
        \coordinate (A) at (0,0);
        \coordinate (B) at (1/2,0);
        \coordinate (C) at (1,0);
        \coordinate (D) at (3/2,0);
        \filldraw[thick, gray!20] (0.25,0) ellipse (0.4cm and 0.17cm);
        \draw[thick, black] (0.25,0) ellipse (0.4cm and 0.17cm);
         \draw[thick] (A) -- (B) -- (C) -- (D);
        \fill[black] (A) circle (2pt);
        \fill[black] (B) circle (2pt);
        \fill[black] (C) circle (2pt);
        \fill[black] (D) circle (2pt);
    \end{tikzpicture}};}\mathcal{R}_{\tikz[baseline,yshift=0.5ex] \node[anchor=base] at (0,0) {\begin{tikzpicture}[scale=0.5]
        \coordinate (A) at (0,0);
        \coordinate (B) at (1/2,0);
        \coordinate (C) at (1,0);
        \coordinate (D) at (3/2,0);
        \fillroundrectangle{0}{0.8}{0}{2.2}{0.62}{0};
         \roundrectangle{0}{0.8}{0}{2.2}{0.62}{0};
        \filldraw[thick, white] (0.25,0) ellipse (0.4cm and 0.17cm);
        \draw[thick, black] (0.25,0) ellipse (0.4cm and 0.17cm);
         \draw[thick] (A) -- (B) -- (C) -- (D);
        \fill[black] (A) circle (2pt);
        \fill[black] (B) circle (2pt);
        \fill[black] (C) circle (2pt);
        \fill[black] (D) circle (2pt);
    \end{tikzpicture}};}}, \notag \\
    \frac{1}{Y_{\tikz[baseline,yshift=0.5ex] \node[anchor=base] at (0,0) {\begin{tikzpicture}[scale=0.5]
        \coordinate (A) at (0,0);
        \coordinate (B) at (1/2,0);
        \coordinate (C) at (1,0);
        \coordinate (D) at (3/2,0);
        \fillroundrectangle{0}{0.8}{0}{2.2}{0.62}{0};
         \roundrectangle{0}{0.8}{0}{2.2}{0.62}{0};
        \filldraw[thick, red!20] (0.25,0) ellipse (0.5cm and 0.22cm);
        \draw[thick, black] (0.25,0) ellipse (0.5cm and 0.22cm);
        \draw[black,thick] (A) circle (4pt);
        \draw[thick] (A) -- (B) -- (C) -- (D);
        \fill[black] (A) circle (2pt);
        \fill[black] (B) circle (2pt);
        \fill[black] (C) circle (2pt);
        \fill[black] (D) circle (2pt);
    \end{tikzpicture}};}} &= \frac{1}{\mathcal{R}_{\tikz[baseline,yshift=0.5ex] \node[anchor=base] at (0,0) {\begin{tikzpicture}[scale=0.5]
        \coordinate (A) at (0,0);
        \coordinate (B) at (1/2,0);
        \coordinate (C) at (1,0);
        \coordinate (D) at (3/2,0);
         \draw[thick] (A) -- (B) -- (C) -- (D);
        \filldraw[gray!20] (A) circle (4pt);
        \draw[black,thick] (A) circle (4pt);
        \fill[black] (A) circle (2pt);
        \fill[black] (B) circle (2pt);
        \fill[black] (C) circle (2pt);
        \fill[black] (D) circle (2pt);
    \end{tikzpicture}};}\mathcal{R}_{\tikz[baseline,yshift=0.5ex] \node[anchor=base] at (0,0) {\begin{tikzpicture}[scale=0.5]
        \coordinate (A) at (0,0);
        \coordinate (B) at (1/2,0);
        \coordinate (C) at (1,0);
        \coordinate (D) at (3/2,0);
        \filldraw[thick, gray!20] (0.25,0) ellipse (0.5cm and 0.22cm);
        \draw[thick, black] (0.25,0) ellipse (0.5cm and 0.22cm);
         \draw[thick] (A) -- (B) -- (C) -- (D);
        \filldraw[white] (A) circle (4pt);
        \draw[black,thick] (A) circle (4pt);
        \fill[black] (A) circle (2pt);
        \fill[black] (B) circle (2pt);
        \fill[black] (C) circle (2pt);
        \fill[black] (D) circle (2pt);
    \end{tikzpicture}};}}+\frac{1}{\mathcal{R}_{\tikz[baseline,yshift=0.5ex] \node[anchor=base] at (0,0) {\begin{tikzpicture}[scale=0.5]
        \coordinate (A) at (0,0);
        \coordinate (B) at (1/2,0);
        \coordinate (C) at (1,0);
        \coordinate (D) at (3/2,0);
        \filldraw[thick, gray!20] (0.25,0) ellipse (0.5cm and 0.22cm);
        \draw[thick, black] (0.25,0) ellipse (0.5cm and 0.22cm);
         \draw[thick] (A) -- (B) -- (C) -- (D);
        \filldraw[white] (A) circle (4pt);
        \draw[black,thick] (A) circle (4pt);
        \fill[black] (A) circle (2pt);
        \fill[black] (B) circle (2pt);
        \fill[black] (C) circle (2pt);
        \fill[black] (D) circle (2pt);
    \end{tikzpicture}};}\mathcal{R}_{\tikz[baseline,yshift=0.5ex] \node[anchor=base] at (0,0) {\begin{tikzpicture}[scale=0.5]
        \coordinate (A) at (0,0);
        \coordinate (B) at (1/2,0);
        \coordinate (C) at (1,0);
        \coordinate (D) at (3/2,0);
        \fillroundrectangle{0}{0.8}{0}{2.2}{0.62}{0};
         \roundrectangle{0}{0.8}{0}{2.2}{0.62}{0};
        \filldraw[thick, white] (0.25,0) ellipse (0.4cm and 0.17cm);
        \draw[thick, black] (0.25,0) ellipse (0.4cm and 0.17cm);
         \draw[thick] (A) -- (B) -- (C) -- (D);
        \fill[black] (A) circle (2pt);
        \fill[black] (B) circle (2pt);
        \fill[black] (C) circle (2pt);
        \fill[black] (D) circle (2pt);
    \end{tikzpicture}};}},
\end{align}
and keep only terms with simple poles. The contribution to the canonical form \eqref{eq:CP4_vertex} from this vertex provides the correct contribution to the cosmological amplitube given by
\begin{align}
\Psi_{P_4} = \frac{1}{\mathcal{R}_{\tikz[baseline,yshift=0.5ex] \node[anchor=base] at (0,0) {\begin{tikzpicture}[scale=0.5]
        \coordinate (A) at (0,0);
        \coordinate (B) at (1/2,0);
        \coordinate (C) at (1,0);
        \coordinate (D) at (3/2,0);
         \draw[thick] (A) -- (B) -- (C) -- (D);
        \filldraw[gray!20] (A) circle (4pt);
        \draw[black,thick] (A) circle (4pt);
        \fill[black] (A) circle (2pt);
        \fill[black] (B) circle (2pt);
        \fill[black] (C) circle (2pt);
        \fill[black] (D) circle (2pt);
    \end{tikzpicture}};}\mathcal{R}_{\tikz[baseline,yshift=0.5ex] \node[anchor=base] at (0,0) {\begin{tikzpicture}[scale=0.5]
        \coordinate (A) at (0,0);
        \coordinate (B) at (1/2,0);
        \coordinate (C) at (1,0);
        \coordinate (D) at (3/2,0);
         \draw[thick] (A) -- (B) -- (C) -- (D);
        \filldraw[gray!20] (D) circle (4pt);
        \draw[black,thick] (D) circle (4pt);
        \fill[black] (A) circle (2pt);
        \fill[black] (B) circle (2pt);
        \fill[black] (C) circle (2pt);
        \fill[black] (D) circle (2pt);
    \end{tikzpicture}};}\mathcal{R}_{\tikz[baseline,yshift=0.5ex] \node[anchor=base] at (0,0) {\begin{tikzpicture}[scale=0.5]
        \coordinate (A) at (0,0);
        \coordinate (B) at (1/2,0);
        \coordinate (C) at (1,0);
        \coordinate (D) at (3/2,0);
        \filldraw[thick, gray!20] (0.25,0) ellipse (0.4cm and 0.17cm);
        \draw[thick, black] (0.25,0) ellipse (0.4cm and 0.17cm);
         \draw[thick] (A) -- (B) -- (C) -- (D);
        \fill[black] (A) circle (2pt);
        \fill[black] (B) circle (2pt);
        \fill[black] (C) circle (2pt);
        \fill[black] (D) circle (2pt);
    \end{tikzpicture}};}\mathcal{R}_{\tikz[baseline,yshift=0.5ex] \node[anchor=base] at (0,0) {\begin{tikzpicture}[scale=0.5]
        \coordinate (A) at (0,0);
        \coordinate (B) at (1/2,0);
        \coordinate (C) at (1,0);
        \coordinate (D) at (3/2,0);
        \filldraw[thick, gray!20] (0.25,0) ellipse (0.5cm and 0.22cm);
        \draw[thick, black] (0.25,0) ellipse (0.5cm and 0.22cm);
         \draw[thick] (A) -- (B) -- (C) -- (D);
        \filldraw[white] (A) circle (4pt);
        \draw[black,thick] (A) circle (4pt);
        \fill[black] (A) circle (2pt);
        \fill[black] (B) circle (2pt);
        \fill[black] (C) circle (2pt);
        \fill[black] (D) circle (2pt);
    \end{tikzpicture}};}\mathcal{R}_{\tikz[baseline,yshift=0.5ex] \node[anchor=base] at (0,0) {\begin{tikzpicture}[scale=0.5]
        \coordinate (A) at (0,0);
        \coordinate (B) at (1/2,0);
        \coordinate (C) at (1,0);
        \coordinate (D) at (3/2,0);
        \fillroundrectangle{0}{0.8}{0}{2.2}{0.62}{0};
         \roundrectangle{0}{0.8}{0}{2.2}{0.62}{0};
        \filldraw[thick, white] (.25,0) ellipse (0.45cm and 0.2cm);
        \draw[thick, black] (.25,0) ellipse (0.45cm and 0.2cm);
         \draw[thick] (A) -- (B) -- (C) -- (D);
        \filldraw[white] (D) circle (4pt);
        \draw[black,thick] (D) circle (4pt);
        \fill[black] (A) circle (2pt);
        \fill[black] (B) circle (2pt);
        \fill[black] (C) circle (2pt);
        \fill[black] (D) circle (2pt);
    \end{tikzpicture}};}\mathcal{R}_{\tikz[baseline,yshift=0.5ex] \node[anchor=base] at (0,0) {\begin{tikzpicture}[scale=0.5]
        \coordinate (A) at (0,0);
        \coordinate (B) at (1/2,0);
        \coordinate (C) at (1,0);
        \coordinate (D) at (3/2,0);
        \fillroundrectangle{0}{0.8}{0}{2.2}{0.62}{0};
         \roundrectangle{0}{0.8}{0}{2.2}{0.62}{0};
        \filldraw[thick, white] (0.25,0) ellipse (0.4cm and 0.17cm);
        \draw[thick, black] (0.25,0) ellipse (0.4cm and 0.17cm);
         \draw[thick] (A) -- (B) -- (C) -- (D);
        \fill[black] (A) circle (2pt);
        \fill[black] (B) circle (2pt);
        \fill[black] (C) circle (2pt);
        \fill[black] (D) circle (2pt);
    \end{tikzpicture}};}} + \ldots.
\end{align}

\section{Outlook and Conclusions}\label{sec:outlook}
In this paper we introduced a new class of polytopes, that we call the graph cosmohedra, that are of interest to mathematicians and physicists alike. From the physics point of view, they generalise the recently introduced cosmohedra that capture the combinatorics of singularities of the wavefunctions for $\text{tr}(\phi^3)$ theory. For mathematicians, we provide a combinatorial and geometric realisation of a new class of polytope that generalise associahedra and permutohedra. The combinatorial stratification of the graph cosmohedra is given by regional tubings that generalise polyangulations of $n$-gons. We also provide an explicit embedding of these polytopes \eqref{eq_ABHY_cosmo} that generalises the ABHY-like embedding \eqref{eq_ABHY_assoc} initially studied for associahedra. To each graph associahedron, we associate a function -- amplitube -- that possesses interesting factorisation properties reminiscent of the factorisations of scattering amplitudes in high energy physics. Whilst, for each graph cosmohedron, we associate a function -- cosmological amplitube -- that possesses factorisation properties reminiscent of the cosmological wavefunctions of $\tr(\phi^3)$ theory. 

There are a number of future research direction which our work suggests.
The graph associahedra studied here are a special case of {\it generalised permutohedra} introduced and studied in \cite{postnikov2009permutohedra,feichtner2004matroid,dovsen2011hypergraph}. As such it would be interesting to extend the construction presented here to generalised permutohedra in order to define their corresponding `cosmological' polytopes. We expect the resulting construction to be closely related to the work of \cite{gaiffi2015permutonestohedra}. In a similar direction 
the CEGM amplitudes introduced in \cite{Cachazo:2019ngv} provide a vast generalisation of the familiar $\tr (\phi^3)$ amplitudes, and it would be interesting to see whether a cosmological version of these generalised amplitudes exists as well.

The authors of \cite{Arkani-Hamed:2024jbp} introduced yet another new family of polytopes referred to as {\it correlahedra}, which encode the geometry of cosmological correlation functions. The correlahedra have two special facets, one which takes the form of the associahedron, and one which takes the form of the cosmohedron. As such the correlahedra live in one higher dimension than these polytopes. It is natural to extend the notion of correlahedra to {\it graph correlahedra} which have two special facets corresponding to the graph associahedron and graph cosmohedron. Since the embedding of the graph associahedra/cosmohedra presented here are already described in a one higher dimensional space we expect a simple extension of our inequalities to recover the graph correlahedron.

Finally, in the case of the graph $P_{n+1}$, the regional tubings introduced here provide a way to encode all sub-polygons of the $n$-gon on a single graph. Since the recently discovered {\it kinematic flow}, which governs the structure of differential equations satisfied by FRW correlators \cite{Arkani-Hamed:2023kig,Arkani-Hamed:2023bsv}, is naturally phrased in terms of sub-polygons, it would be interesting to see whether the regional tubings can be generalised to also describe the kinematic flow.

 \section*{Acknowledgements}
RG would like to thank Stefan Forcey and Satyan Devadoss for comments on how the construction in this paper fits into the mathematics literature, in particular for bringing to our attention the references \cite{kapranov1993permutoassociahedron,gaiffi2015permutonestohedra}.
\bibliographystyle{nb}

\bibliography{graph_cosmohedron}

\begin{thebibliography}{10}
\ifx\href\asklfhas\newcommand{\href}[2]{#2}\fi
\ifx\arxivref\asklfhas\newcommand{\arxivref}[2]{\href{http://arxiv.org/abs/#1}{#2}}\fi
\ifx\doiref\asklfhas\newcommand{\doiref}[2]{\href{http://dx.doi.org/#1}{#2}}\fi
\raggedright
\small
\parskip 0pt

\bibitem{Arkani-Hamed:2017mur}
N.~Arkani-Hamed, Y.~Bai, S.~He and G.~Yan,
\textit{``{Scattering Forms and the Positive Geometry of Kinematics, Color and
  the Worldsheet}''},
\textsf{\doiref{10.1007/JHEP05(2018)096}{JHEP~1805,~096~(2018)}},
\texttt{\arxivref{1711.09102}{arxiv:1711.09102}}.

\bibitem{Arkani-Hamed:2017tmz}
N.~Arkani-Hamed, Y.~Bai and T.~Lam,
\textit{``{Positive Geometries and Canonical Forms}''},
\textsf{\doiref{10.1007/JHEP11(2017)039}{JHEP~1711,~039~(2017)}},
\texttt{\arxivref{1703.04541}{arxiv:1703.04541}}.

\bibitem{Arkani-Hamed:2024jbp}
N.~Arkani-Hamed, C.~Figueiredo and F.~Vaz\~ao,
\textit{``{Cosmohedra}''},
\texttt{\arxivref{2412.19881}{arxiv:2412.19881}}.

\bibitem{Arkani-Hamed:2023kig}
N.~Arkani-Hamed, D.~Baumann, A.~Hillman, A.~Joyce, H.~Lee and G.~L.~Pimentel,
\textit{``{Differential Equations for Cosmological Correlators}''},
\texttt{\arxivref{2312.05303}{arxiv:2312.05303}}.

\bibitem{Arkani-Hamed:2023bsv}
N.~Arkani-Hamed, D.~Baumann, A.~Hillman, A.~Joyce, H.~Lee and G.~L.~Pimentel,
\textit{``{Kinematic Flow and the Emergence of Time}''},
\texttt{\arxivref{2312.05300}{arxiv:2312.05300}}.

\bibitem{Arkani-Hamed:2017fdk}
N.~Arkani-Hamed, P.~Benincasa and A.~Postnikov,
\textit{``{Cosmological Polytopes and the Wavefunction of the Universe}''},
\texttt{\arxivref{1709.02813}{arxiv:1709.02813}}.

\bibitem{De:2023xue}
S.~De and A.~Pokraka,
\textit{``{Cosmology meets cohomology}''},
\textsf{\doiref{10.1007/JHEP03(2024)156}{JHEP~2403,~156~(2024)}},
\texttt{\arxivref{2308.03753}{arxiv:2308.03753}}.

\bibitem{Benincasa:2024leu}
P.~Benincasa and G.~Dian,
\textit{``{The Geometry of Cosmological Correlators}''},
\texttt{\arxivref{2401.05207}{arxiv:2401.05207}}.

\bibitem{Benincasa:2024lxe}
P.~Benincasa and F.~Vaz\~ao,
\textit{``{The Asymptotic Structure of Cosmological Integrals}''},
\texttt{\arxivref{2402.06558}{arxiv:2402.06558}}.

\bibitem{De:2024zic}
S.~De and A.~Pokraka,
\textit{``{A physical basis for cosmological correlators from cuts}''},
\texttt{\arxivref{2411.09695}{arxiv:2411.09695}}.

\bibitem{Cachazo:2019ngv}
F.~Cachazo, N.~Early, A.~Guevara and S.~Mizera,
\textit{``{Scattering Equations: From Projective Spaces to Tropical
  Grassmannians}''},
\textsf{\doiref{10.1007/JHEP06(2019)039}{JHEP~1906,~039~(2019)}},
\texttt{\arxivref{1903.08904}{arxiv:1903.08904}}.

\bibitem{Lam:2024jly}
T.~Lam,
\textit{``{Matroids and amplitudes}''},
\texttt{\arxivref{2412.06705}{arxiv:2412.06705}}.

\bibitem{Arkani-Hamed:2024pzc}
N.~Arkani-Hamed, H.~Frost and G.~Salvatori,
\textit{``{The Cut Equation}''},
\texttt{\arxivref{2412.21027}{arxiv:2412.21027}}.

\bibitem{carr2006coxeter}
M.~Carr and S.~L.~Devadoss,
\textit{``Coxeter complexes and graph-associahedra''},
\textsf{Topology~and~its~Applications~153,~2155~(2006)}.

\bibitem{kapranov1993permutoassociahedron}
M.~M.~Kapranov,
\textit{``The permutoassociahedron, Mac Lane's coherence theorem and asymptotic
  zones for the KZ equation''},
\textsf{Journal~of~Pure~and~Applied~Algebra~85,~119~(1993)}.

\bibitem{devadoss2009realization}
S.~L.~Devadoss,
\textit{``A realization of graph associahedra''},
\textsf{Discrete~Mathematics~309,~271~(2009)}.

\bibitem{He:2020onr}
S.~He, Z.~Li, P.~Raman and C.~Zhang,
\textit{``{Stringy canonical forms and binary geometries from associahedra,
  cyclohedra and generalized permutohedra}''},
\textsf{\doiref{10.1007/JHEP10(2020)054}{JHEP~2010,~054~(2020)}},
\texttt{\arxivref{2005.07395}{arxiv:2005.07395}}.

\bibitem{postnikov2009permutohedra}
A.~Postnikov,
\textit{``Permutohedra, associahedra, and beyond''},
\textsf{International~Mathematics~Research~Notices~2009,~1026~(2009)}.

\bibitem{Brown:2025jjg}
F.~Brown and C.~Dupont,
\textit{``{Positive geometries and canonical forms via mixed Hodge theory}''},
\texttt{\arxivref{2501.03202}{arxiv:2501.03202}}.

\bibitem{feichtner2004matroid}
E.~M.~Feichtner and B.~Sturmfels,
\textit{``Matroid polytopes, nested sets and Bergman fans''},
\textsf{arXiv~preprint~math/0411260~2009,~B.~Sturmfels~(2004)}.

\bibitem{dovsen2011hypergraph}
K.~Do{\v{s}}en and Z.~Petri{\'c},
\textit{``Hypergraph polytopes''},
\textsf{Topology~and~its~Applications~158,~1405~(2011)}.

\bibitem{gaiffi2015permutonestohedra}
G.~Gaiffi,
\textit{``Permutonestohedra''},
\textsf{Journal~of~Algebraic~Combinatorics~41,~125~(2015)}.

\end{thebibliography}

\end{document}